\documentclass[aas_macros]{cplslarge}%

\usepackage{color}
\usepackage{xspace}
\usepackage{graphicx}
\usepackage{amssymb}
\usepackage{amsmath}
\usepackage[authoryear]{natbib}

\usepackage{makeidx}

\include{mnras_jdf}
\begin{document}
\mainmatter
\author[Latter, Ogilvie, \&\ Rein]{H. N. Latter, G. I. Ogilvie, H. Rein}

\chapter{Planetary rings and other astrophysical disks}

\abstract{This chapter explores the physics shared by
  planetary rings and the various 
  disks that populate the Universe. 
  It begins with an observational overview, ranging from
  protoplanetary disks to spiral galaxies, and then
  compares and contrasts these astrophysical 
  disks with the rings of the Solar System. Emphasis is placed on
  fundamental physics and dynamics, and how
  research into the two classes of object connects. 
  Topics covered include disk formation, accretion, collisional
  processes, waves, instabilities, and satellite-disk interactions.}

\section{INTRODUCTION}

Disks are ubiquitous in astrophysics and participate in some of its
most important processes.  Most, but not all,
feed a central mass: by facilitating the transfer of angular momentum,
they permit the accretion of material that would otherwise remain in
orbit \citep{LBPringle74}. As a consequence, disks are essential
to star, planet, and satellite formation
\citep{McKeeOstriker07,WilliamsCieza11,PapTerq06,Peale99}.
They also regulate the growth of supermassive black holes and
thus indirectly influence galactic structure and the intracluster medium
\citep{Vol10,Fabian12}. Although
astrophysical disks can vary by ten orders of magnitude in size
and differ hugely in composition, all share
the same basic dynamics and many physical
phenomena. This review explores these areas of overlap.

The prevalence of flattened astrophysical systems is a result
of dissipation and rotation \citep{GT82}. A cloud of gas or debris
in orbit around a central mass conserves its total angular momentum
but not its energy, as there are
numerous processes that may cool the cloud
(inelastic physical collisions,
Bremsstrahlung, molecular line emission, etc.). 
As a result,
 particles' random velocities are
steadily depleted --- where `random velocity' is understood to be
 the component surplus to
the circular orbit fixed by the angular momentum.
The system
contracts into a flat circular disk, the lowest energy
state accessible. The contraction ends, and an
equilibrium balance achieved, once the cooling is met by heating
(supplied by external irradiation or an internal viscous stress).

Let us define a cylindrical coordinate system with its origin at the
central mass and the vertical pointing in the direction of the total
angular momentum vector. We describe systems as \emph{cold}
when the pressure gradient is weak
and the final equilibrium very thin: radially the
dominant force balance is between the centrifugal force and
gravity, while vertically it is between pressure and gravity.  Systems
described as \emph{hot} have stronger pressure and
settle into thick disks or tori. 
At the far extreme, when rotation is subdominant,
spheroidal morphologies ensue: examples include planets, stars,
globular clusters, elliptical galaxies, etc.

What determines the importance of pressure, and which state is
ultimately achieved, is the relative efficiency of heating and
cooling. In dense planetary rings, energetic losses from strongly
inelastic collisions predominate. Rings are hence
exceptionally cold and thin \citep{Colwelletal09}. In comparison,
radiative cooling in gaseous disks varies by orders of magnitude,
depending on the temperature, dust levels, ionisation fraction, or other
properties \citep[e.g.][]{BellLin94,AF2013}. The heating rate
 also varies considerably,
especially if turbulence or external radiation is present. 
Consequently, gaseous
accretion disks can be thin (though never as thin as dense rings) or
so thick they resemble doughnuts more than they do sheets.

This fundamental paradigm accommodates a diversity of different
astrophysical disk systems, ranging over an enormous variety of length scales,
physical properties, and compositions. In the next section we review
the observational literature on these systems. We then
make clear their key distinguishing physics and physical scales,
extending the schematic account above. The rest of the
chapter visits an assortment of topics that provide enlightening
comparisons between planetary rings and other astrophysical disks. In
particular, we dwell on instances of pollinisation between the
two fields of study. The topics covered include: formation, accretion,
collisional dynamics, waves, instabilities, and finally
satellite--disk interactions. We conclude by speculating on
further connections between planetary rings and other disks that
future work might explore.

\section{OVERVIEW OF ASTROPHYSICAL DISKS}
\label{s:survey}

\subsection{Protoplanetary disks}
\label{s:PPdisks}
\begin{figure}
\figurebox{17pc}{}{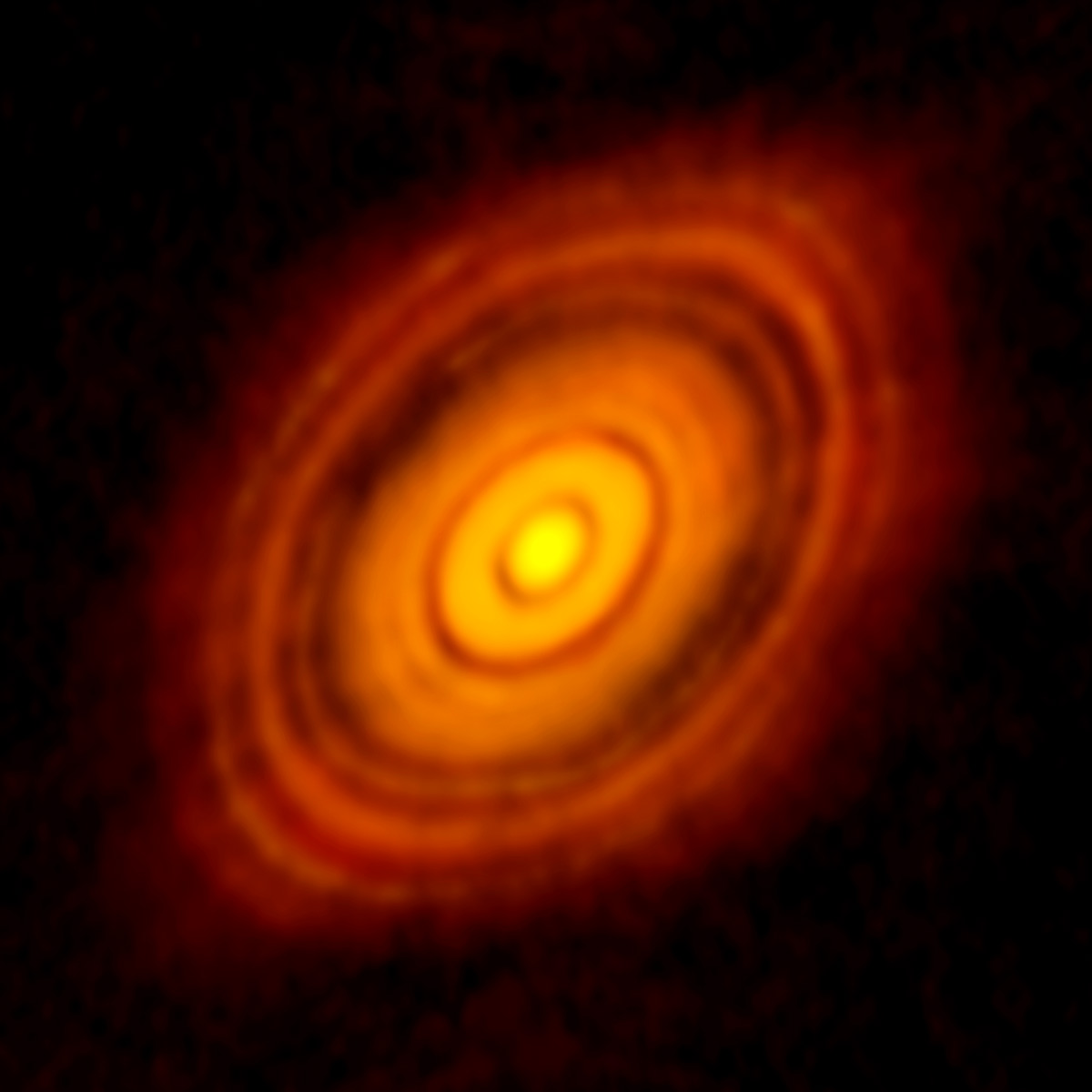}
\caption{ALMA image of the young star HL Tau and its protoplanetary
  disk in the mm continuum.  Credit: ALMA/ESO\label{fig:hltau}}
\end{figure}

Since the Copernican revolution astronomers have recognised that the
planets of the Solar System all orbit the Sun in the same sense, and
almost in the same plane. In the eighteenth century Swedenborg, Kant,
and Laplace, recognising that this arrangement could not have arisen
by chance, proposed that the planets condensed out of a flattened
cloud of gas rotating around the Sun earlier in its life
\citep[][and references therein]{Montmerle2006}.  Their models of
the solar nebula introduced the concept of the protoplanetary disk
(hereafter `PP disk') which, though abandoned briefly in the early
twentieth century \citep[e.g.][]{Jeans17}, lies at the heart of
modern theories
of star and planet formation.

Originally inferred
from the infrared excesses of young stellar objects
\citep[e.g.][]{LadaWilking84}, PP disks were directly imaged first in
the sub-mm \citep{Sarg87,Koerner93}, and then spectacularly in
the optical, when the Hubble
Space Telescope uncovered a number of examples in the Orion
Nebula \citep{McCaughOdell96}. PP disks consist of relatively cool
gas, mostly H$_2$, scattered with dust. Temperatures are $\sim
100$~K generally, but can reach $\sim 3000$~K in the inner
radii. They are believed to survive for a few million years, extend
radially up to $\sim1000$~AU, and exhibit aspect ratios of roughly
$H/r \sim 0.05$, where $H$ is the disk's vertical pressure scaleheight and $r$
is radius \citep{KenHart95,Evans09,WilliamsCieza11}.

Observations of UV excess in the stellar spectrum allow astronomers to estimate the mass
transfer rate through the disk and onto the young star. Comparison of
different systems suggests that accretion is irregular, with
some 50\% of the disk mass falling upon its protostar during less than
10\% of the disk's lifetime \citep{Evans09}. Archetypal systems that
exhibit
fast accretion events are the FU Orionis and EX Lupi variables (FUors
and EXors), which
undergo irregular outbursts of accretion on a range of
long timescales, 50--1000 years \citep{Herbig77,Herbig89,HartKen96,Audard2014}.

\begin{figure}[ht]
\figurebox{16pc}{}{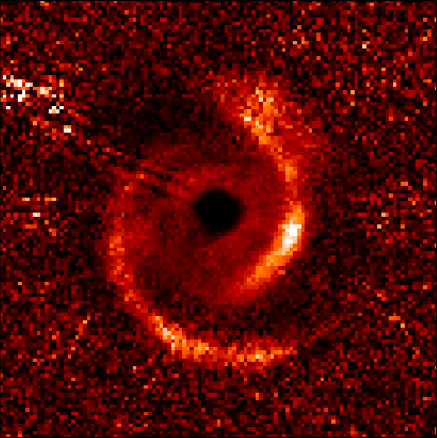}
\caption{The protoplanetary disk MWC 758 as mapped in polarised
  scattered infrared with VLT/SPHERE.
  Credit: \cite{Benisty2015}. \label{fig:MWC758}}
\end{figure}

More recent observations using infrared and radio wavelengths (e.g.\
with Subaru, VLT and ALMA) reveal that PP disks are highly
structured and exhibit gaps, asymmetries, and spirals
\citep{Andrews11,Muto12,Perez14,Brogan15}.  Figure \ref{fig:hltau}
shows an early ALMA image of the disk around HL Tauri, a young
Sun-like star, exhibiting a striking array of rings. Figure
\ref{fig:MWC758} presents a clear example of a spiral density wave in
MWC 758. There is considerable research activity aiming
to explain the features seen in these and similar
images. Embedded planets are the focus of the most popular ideas
\citep[e.g.][]{Tamayo2015}, as theory predicts that they naturally
carve gaps and excite spiral waves \citep[][see also Section
\ref{SDinteractions}]
{PapTerq06}.  However, a panoply of alternatives
 have been proposed that may bear on the observed structures.
These include vortices \citep{VarnTagg06,LesPap10},
gravitational instabilities \citep{Durisen07,TakaInut14}, snow
lines \citep{Zhang2015}, stellar flybys \citep{CP93,X16}, and warps \citep{MPC15}.

\subsection{Dwarf novae, X-ray binaries and Be stars}
\label{s:WDs}

Most stars form in binary systems. 
The more massive (primary)
star evolves faster than its companion (the secondary) and thus ends up
a white dwarf, a neutron
star or a black hole 
while its secondary is left behind on the main sequence.
If the binary orbit is
sufficiently close, the secondary overflows its
critical equipotential surface, or Roche lobe, and spills
over towards the compact primary. Owing to its rotation in the
binary orbit, the transferred gas has too much angular momentum to
fall directly on to the surface of the primary. Instead, it forms an
accretion disk around it. Under the action of torques within the disk,
the gas gradually loses angular momentum, spirals inwards, and is
accreted. 
As the gas falls deeper into the
potential well energy is liberated, and the system becomes luminous
\citep{Warner95,Hellier01}.

\begin{figure}[ht]
\figurebox{16pc}{10pc}{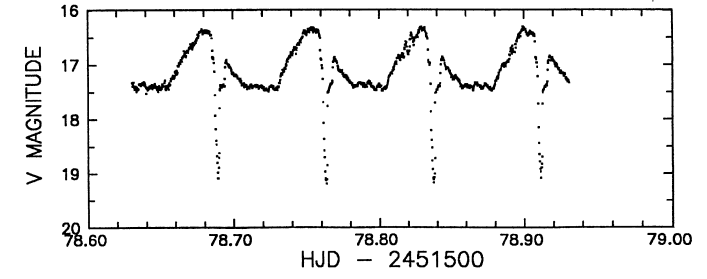}
\caption{Light curve of the dwarf nova IY UMa over roughly four binary
  orbits ($\sim$ 7 hours in total). 
  The vertical axis is magnitude in the V band, and the
  horizontal axis is heliocentric Julian day (HJD).
  The broad humps correspond to the progress of the hot spot around the
  disk, 
  while the abrupt
  dips correspond to eclipses of the white dwarf by the
  secondary. The points are plotted at 17 s intervals. Credit:
  \citet{Pat2000}.}
\label{fig:DNe1}
\end{figure}

Because such disks possess radii similar to that of the Sun, 
they are impossible to resolve directly.
Instead the above picture of disk formation and accretion was deduced 
from detailed analyses of spectra and the peculiarities of
the systems' light curves, which on the orbital timescale ($\sim 1$
hour) exhibit characteristic dips and other features
\citep[e.g.][]{Kraft62,Kraft64,WarnNath71,Smak71}.  
In Figure \ref{fig:DNe1} a representative light curve is
presented, showing two key signatures: 
the eclipse of the compact primary (a white dwarf) and the eclipse of the `hot
spot', the region of the disk struck by the accretion stream from the
secondary star.

\begin{figure}[ht]
\figurebox{18pc}{20pc}{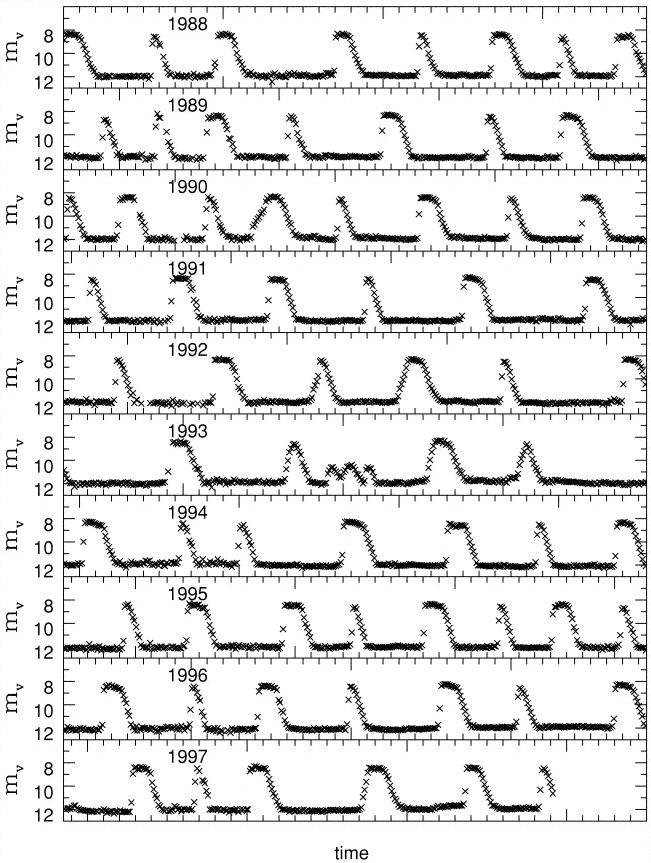}
\caption{Light curve of SS Cyg on long
  timescales, showing quasi-periodic outburst behaviour. Each panel
  represents one year, each small tic on the horizontal axis
  10 days, and each
  cross the daily mean in magnitude --- thus small and fine-scale
  variations (such as in Fig.~\ref{fig:DNe1}) are removed. 
   The first panel begins at HJD 
  2,446,432, and
 the last panel ends at 2,450,082. Credit: \citet{Cannizzo98}.}
\label{fig:DNe2}
\end{figure}

Systems with white dwarf primaries are known as cataclysmic variables
because many of them exhibit dramatic outbursts. These include the
classical novae, in which a layer accreted on the primary ignites in a
thermonuclear runaway \citep{GallStarr78,Shara89}, and the dwarf
novae, in which outbursts occur cyclically and arise from a state
change in the disk itself \citep{Warner95, Lasota01}.  The latter
outbursts feature an increase of brightness of some 2--5 magnitudes,
last for a few days, and possess a recurrence interval of days to
weeks. Figure \ref{fig:DNe2} illustrates a clear sequence of dwarf
nova outbursts in SS Cygni. Note that some sources exhibit more
complicated behaviour such as `standstills' (Z Camelopardalis
variables) and `superoutbursts' (SU Ursae Majoris variables)
\citep{Warner95}.  

Systems with neutron star or black hole primaries are known as X-ray
binaries because their emission is dominated by high-energy photons. 
They were
first discovered by X-ray satellites launched in the 1960s
\citep{Giacconi62,Gursky66,Sandage66}.  In the following 50 years,
space-based X-ray observatories, such as \textit{Einstein},
\textit{Chandra}, and \textit{XMM-Newton}, have uncovered
the properties of many such systems, yet because of their
weak emission in the optical they are far less well constrained than
dwarf novae.  Low-mass X-ray binaries, involving low-mass secondaries,
typically accrete by Roche-lobe overflow as described above. In
contrast, the accretion disks in high-mass X-ray binaries typically
capture their gas from the vigorous winds of the high-mass secondary
stars \citep{CharlesCoe06,TauHeuv06}. In both cases the disk
temperature is strongly affected by X-ray irradiation and, as a result, though
they undergo outbursts, their dynamics can differ markedly from dwarf
novae \citep{Lasota01,King06}.

The disks associated with interacting binaries usually consist of
hydrogen and helium in atomic or ionised form.  Dwarf novae disks are vertically
thin ($H/r \sim 0.01$), but the aspect ratio varies between the low and
the high (outbursting) states, which are
characterised by temperatures between $\sim 3000$~K 
and $\sim50,000$~K at the midplane, respectively \citep{Hellier01}. In the inner
regions of X-ray binaries the disk attains enormous temperatures,
greater than $10^6$~K at the disk surface, 
and cycles through a number of
poorly understood spectral states, some of which are accompanied by
jets
of material
launched perpendicular to the disk 
\citep{RemillardMcClintock2006,DGK07}.

Finally, we touch on Be stars. These are rapidly spinning B stars in
which material is periodically expelled from the equator and
forms a centrifugally
supported disk. Usually the disks are inferred observationally by their characteristic
Balmer lines and their polarisation of the star's continuum radiation
\citep{Rivinius13}, though in the case of $\zeta$ Tauri, a map of 
the disk itself has been directly reconstructed using interferometry
in the H$\alpha$ line
\citep{Quirrenbach94}, while peculiarities in the light curves of the
eclipsing binaries $\epsilon$ Aurigae and EE Cephei are best explained by a
tilted disk around a B star \citep{Hoard10,Galan12}.
 The mechanism by which material is ejected
from the star (the `Be phemonenon') is still debated, but may involve
non-radial stellar pulsations, magnetic fields, or winds. Once
formed, the disk evolves viscously, while undergoing global
eccentric oscillations of unknown provenance
\citep{Okazaki91}.

\subsection{Debris disks}
\label{s:Debris}

Debris disks consist of dust and larger solids usually orbiting a
young or main-sequence star \citep{Wyatt08}.  They are the
end-points of protoplanetary disk evolution, the gas either
accreted or swept away by a photoevaporative wind.
Alternatively, they may be regarded as
the `leftovers' of the planet
formation process \citep{Alex06,Wyatt15}.

The first debris disk was discovered orbiting Vega by the
\textit{IRAS} satellite, its presence betrayed by a large infrared
excess \citep{Aum84}.  Thanks to recent surveys by \textit{Spitzer}
and \textit{Herschel}, there are now several hundred debris disk
candidates circling a wide variety of stars \citep{Su06,Eiroa13}. 
 Debris disks have also
been observed in sub-mm to optical wavelengths and may now
 be imaged directly, giving astronomers
information about their morphology \citep{Wyatt08}.  These
disks are believed to be extrasolar analogues of the asteroid belt,
Kuiper belt, and exozodiacal dust in our own Solar System, and so the
study of debris disks straddles the two fields of planetary science
and astrophysics.

\begin{figure}[ht]
\figurebox{20pc}{}{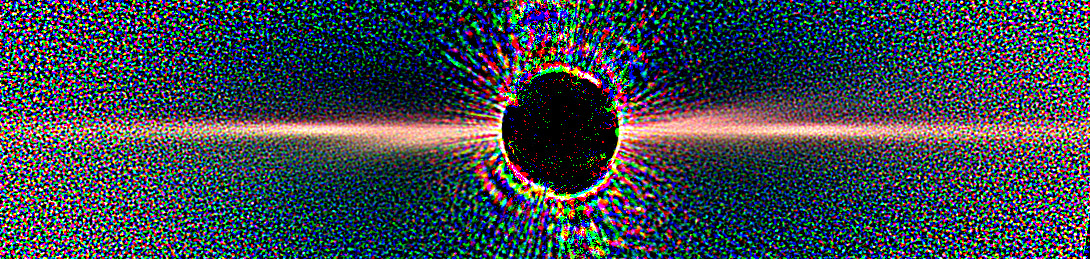}
\caption{This image shows the edge-on disk around Beta Pictoris, taken
  by the Hubble Space Telescope. One can identify a primary disk and a
  secondary, slightly tilted, disk.
 Credit: ESA/Hubble\label{fig:betapic}.}
\end{figure}

The majority of observed debris disk dust lies in the $1$--$100$
$\mu$m size range, and hence
their infrared emission is difficult to study from the
ground. Typically, the total mass in these grains is significantly
less than that of the Earth \citep{Wyatt03,Wyatt08}, and is
distributed between radii $10$--$100$ AU. Rather than being
primordial, the dust must be constantly replenished 
or else it would be eliminated by radiation pressure, collisional
fragmentation, or stellar wind drag. Infrequent impacts between larger
bodies are thought to initiate a collisional cascade that supplies
this material \citep{Backman93,Wyatt02}.
Unfortunately, the population of large solids is difficult to observe, 
owing to their smaller total surface
area. Planetary sized objects, however, have been directly
imaged and also inferred from perturbations in the dust (see
below).  

In comparison, we know of several hundred
bodies with a size of roughly $100$ km in the Kuiper belt, and have
constrained some of their surface properties
\citep{Petit08,Stansberry08}.  Moreover, to explain the frequency of
short-period comets, theoretical estimates show that the belt must contain
at least $10^8$ bodies with sizes greater than $1$ km
\citep{Fari96,Jewitt00}. These estimates provide data on intermediate size
bodies in one
debris disk, at least.

Direct imaging reveals that debris disks exhibit a range of intriguing
morphologies: sharp edges, gaps, warps, rings, spirals, asymmetries,
and clumps \citep{Wyatt08}.  Figure \ref{fig:betapic} shows one of the
nearest debris disks, around the star Beta Pictoris. 
One can identify two
disks slightly tilted with respect to each other.  Planets
can potentially sculpt and structure debris disks
\citep[e.g.][]{Mouillet97,Wyatt05b,Quillen06,Su09}, and indeed the
tilt evident in Figure \ref{fig:betapic} is thought to be forced by 
a massive Jovian planet \citep{Lagrange09}.

\subsection{Active galactic nuclei}

It is generally accepted that most galaxies 
contain a supermassive black hole, of
up to a few billion solar masses, at their centres \citep{Ferrarese05,Merritt13}. A
small proportion of these are `active', in that they produce immense
and persistent volumes of radiation, sometimes orders of magnitude 
greater than the total power output of their host galaxies. The origin of this
spectacular luminosity is the accretion of matter, through
which gravitational energy is converted into
mechanical and electromagnetic energy. 
Because of the intense gravity of supermassive black holes, accretion of only one solar
mass per year is required to generate the observed luminosities
\citep{Salpeter64,LB69,Marconi04}.

\begin{figure}[ht]
\figurebox{20pc}{}{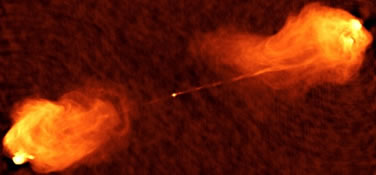}
\caption{A radio map of the galaxy Cygnus A (at a wavelength of 6
  cm). 
   The two jets can be
  seen emerging from the nucleus of the galaxy and colliding with the
  intergalactic medium in the two large radio lobes, each roughly 
  100 kpc from the AGN.  Credit:
  NRAO/AUI. \label{fig:Cygnus A}}
\end{figure}

The spectrum of these `active galactic nuclei' (AGN) is
strikingly broad, and can extend from the far infrared to hard
X-rays; for example, the well studied case of NGC 4151 emits roughly
the same specific intensity over five orders of magnitude in frequency
\citep{Ulrich00}. The optical to extreme ultraviolet light emerges from
the accretion disk surrounding the black hole, while the X-rays are generated
by the disk's hot corona of dilute gas.
Dust in the disk, or in
the systems' enveloping torus, is responsible for the infrared \citep{Ferrarese05}. 
The optical and UV emission tells us that the disk surface is no hotter
than about $10^5$ K, though temperatures at the midplane near the black hole can greatly
exceed that. Variability on short timescales
suggests that disk sizes are typically 1000 AU or less 
\citep[e.g.][]{Green64,Peterson01}. 
In addition, AGN exhibit strong and broad spectral lines: their large
redshifts
reveal their distance from our galaxy, while their Doppler-broadened linewidths
can constrain the mass of the central black hole, as in the case of
M87 \citep{Macchetto97}.
Finally, a subclass of AGN is characterised by powerful radio emission
\citep[e.g.][]{Baade54,Fanaroff74,Miley08}, usually
accompanied by non-thermal gamma rays \citep{Hart99}. 
This radiation issues from intense
relativistic jets oriented perpendicular to the disk plane
\citep[e.g.][]{Biretta99}. Figure \ref{fig:Cygnus A} shows a radio
image of such a jet emerging from Cygnus A.

Owing to this collection of varied and unusual properties, 
several decades passed before AGN were properly 
identified and understood. In fact, for some
time they were classified as separate and distinct sources:
Seyfert galaxies, radio galaxies, BL Lac objects, quasars, and
blazars. 
By the early 1980s it was becoming clearer that these classes
were different
`faces' of the same astrophysical object, an accreting supermassive
black hole, with the variation in
their observed properties attributed mainly to differing viewing angles, 
and the presence, or not, of a jet  \citep{Rees84,Antonucci93}. See \citet{Netzer2015} for a
recent discussion of AGN `unification' theories.

AGN are of immense importance in galactic astronomy and cosmology.
They impact on the structure and evolution of their host
galaxies, clearly demonstrated by the strong correlation between an AGN's
 mass and the velocity dispersion of its host galaxy's stars, on one hand,
 and on the size of the galactic central bulge, on the other 
 \citep{Kormendy95,McConnell11}. They control, to some extent, the
 dynamics of the intracluster medium of galaxies, via the
deposition of mechanical energy through jets (`AGN feedback'),
a process that bears directly on the cooling-flow problem in such
systems \citep{Fabian12}. AGN emission may also act as a probe of the
intervening gas between our galaxy and high redshifts 
\citep[e.g.][]{Fan06}.
At the same time they pose a number of challenging problems:
how do the black holes grow so large? How are relativistic jets launched, and why
do only some AGN produce them? 
Why are they more numerous at large redshift, and does this mean that
AGN represent a transient phase of galactic evolution?

\subsection{Tidal disruption events}
\label{s:TDEs}
Spectroscopic evidence indicates that some 20\% of white dwarf
photospheres are polluted with metal elements
\citep[the `DAZ phenomenon',][]{Zuck03}.
Moreover, these stars must be
continually accreting new pollutants because of the short 
sinking time of some observed
ions (e.g.\ Mg II) \citep{Holberg97,Koester97}. 
An additional clue is that a fraction of the most contaminated examples
display evidence of a circumstellar dust 
ring from either infrared excess \citep[e.g][]{Farihi09} or
double-peaked optical emission lines \citep{Gansicke06}.
As a result, astronomers have deduced 
that DAZ white dwarfs are ringed by narrow accretion disks of
debris and gas, probably the result of the tidal disruption of an asteroid or
minor planet \citep{Graham90,Debes02,Jura03,Jura08}.

One problem this scenario must overcome is how to supply the white
dwarf with a suitable body to disrupt. Asteroid belts
and/or 
planets on wide orbits can survive the giant-branch precursor to the white
dwarf \citep{Vill07,Bon10}, 
but their orbits must be subsequently perturbed so that they
plunge to sufficiently small radii. Scattering of asteroids by planets
is the currently favoured model, the planetary system wrought
dynamically unstable in the post-stellar mass-loss stage
\citep{Bon11,Debes12}.
The intense interest driving this field centres on the
make-up of the pollution itself, because it provides an opportunity to
directly sample the compositions of rocky
exoplanetary systems \citep[e.g.][]{Klein10,Gansicke12,Xu14}.
 
Similarly, when stars stray too near supermassive black holes they are
tidally ripped apart. At early times the stellar material falls on the
black hole at a tremendous rate, 
resulting in a flare whose luminosity approaches that of a supernova. 
After this initial period ($\sim 10$ days), the remaining
stellar mass accretes via a narrow disk over the course of a few
years, the emission at this stage peaking in the UV and soft X-rays
\citep{Rees88,Strubbe09}. First hypothesised in the 1970s
\citep{Hills75}, such tidal disruption events were not observed until the
\emph{ROSAT} All-Sky survey 20 years later \citep[see][]{Kom99,Don02}. 
A dozen candidates have been revealed since, some
 accompanied by short-lived relativistic jets
\citep{Bloom11,Burrows11}. 

The astrophysical interest in these impressive events lies
in their ability to identify and characterise 
\emph{quiescent} supermassive black holes, which would otherwise lie undetected 
at the centres of galaxies. They can help astronomers determine
black hole spin and study
jet launching and accretion disk physics; they may also provide an
electromagnetic counterpart to the gravitational waves emitted during the
initial accretion of the star \citep[e.g.][]{Kobayashi04,Kesden12}.

\subsection{Galactic disks}

It was Kant, again, who first proposed that the Milky Way was a
system of stars orbiting collectively in much the same way as the
Solar System. Moreover, he speculated that observed nebulae might
be distant `island
universes', as vast as the Milky Way. This
hypothesis was partly verified by Lord Rosse in 1845, 
who resolved M51, and a number of other sources, into spiral patterns
\citep{BinneyMerri98}. But it was not till Hubble's observations of
Cepheid variable stars in M33 that it was
firmly established that the nebulae were in fact (far) 
outside the Milky Way and were indeed independent `island' galaxies
\citep{Hubble25}.   

\begin{figure}[ht]
\figurebox{20pc}{}{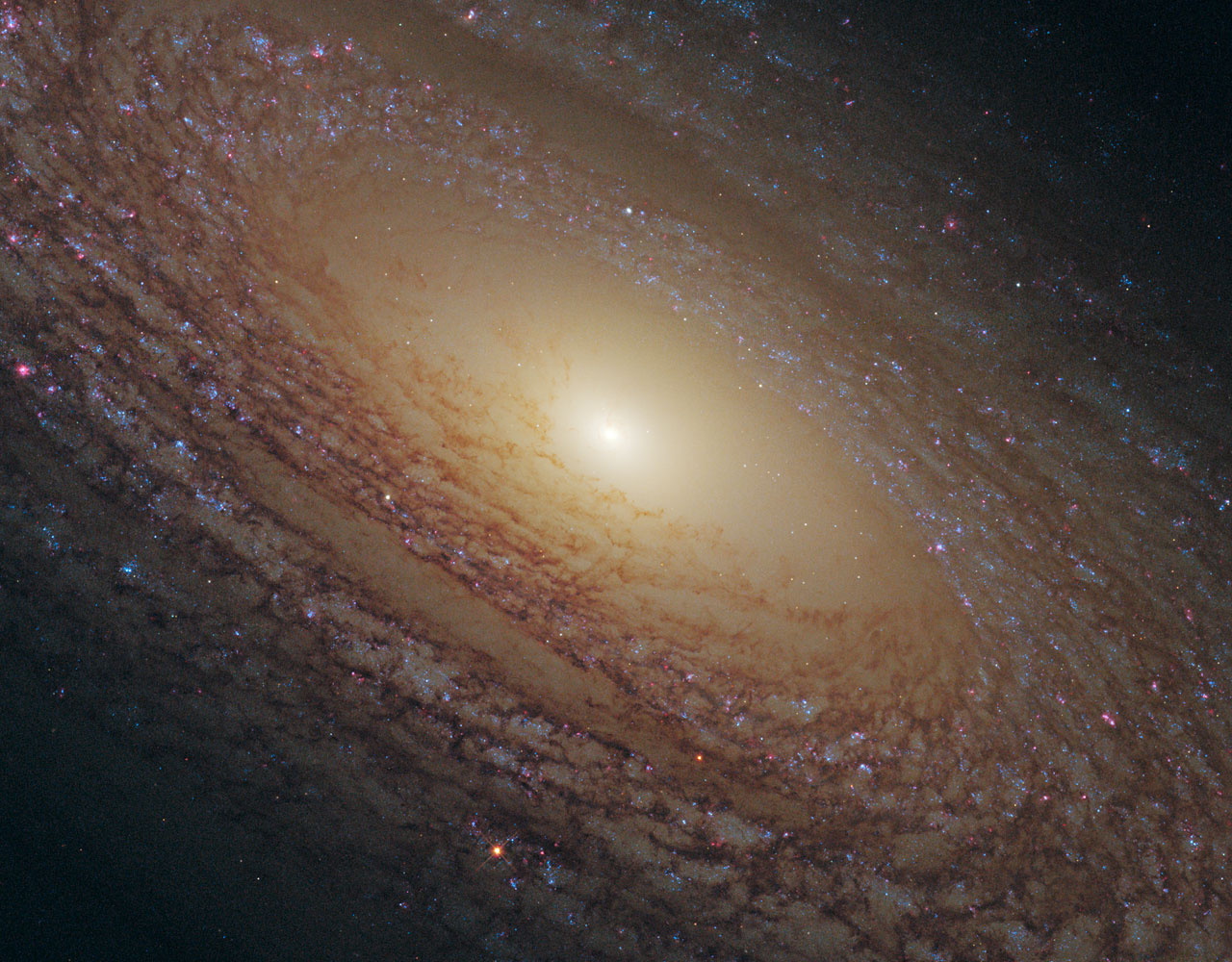}
\caption{An HST image of the flocculent spiral
 galaxy NGC 2841. Credit: ESA/HST. \label{fig:NGC2841}}
\end{figure}

Disk galaxies are flattened structures composed of stars, gas, and dust
undergoing, for the most part, rotational motion around the galactic
nucleus. 
Originally classified by \cite{Hubble1936} 
as part of the Hubble sequence, they differ from most astrophysical disks 
in that their orbital motion
is not entirely dominated by the
central object (a supermassive black hole); 
their rotation curves hence
depart significantly from Keplerian. Aside from the initial stage of galaxy formation,
there is no accretion throughout the whole disk, per se, mainly on
account of there being insufficient rotation periods during a galactic lifetime.
 Most stars are clearly luminous in the
optical and UV; the gas, on the other hand, falls into a variety
of thermodynamic phases that
emit in correspondingly diverse frequency bands 
\citep{Field65,McKee77,Draine11}.

Galactic disks manifest large-scale features such as
spiral arms, central bars, rings, and streams.
They also generally possess a central spheroidal bulge, containing older
stars, and a large spherical dark matter
halo which, 
according to large-scale cosmological simulations, plays an essential
part in
their formation \citep{Springel2005}.
Spiral galaxies may be grouped
into various classes, the distinctions resting on
how tightly wound the spirals are \citep{BinneyMerri98}, and
are sometimes labelled as `grand design' or `flocculent'
depending on the coherence of the spiral arms. Figure
\ref{fig:NGC2841} shows an example of a flocculent spiral galaxy.
Lenticular galaxies, on the other hand, are regarded as intermediate between
elliptical and spiral galaxies. They do not exhibit spiral arms but can form
bars and rings.

\section{SCALES, PARAMETERS, PHYSICAL MODELS}

Disks encompass a vast range of scales and compositions but
they all share the same fundamental force balance:
the centrifugal force matches the radial component of the central object's gravity. 
The mutual cancellation of these powerful forces releases into the
dynamical arena a host of 
subdominant processes that provide the 
inherent variety and interest of astrophysical disks \citep{GorkFrid94}. 
Some of these we cover later in this chapter. 
For now we provide a list of key scales and dimensionless quantities
that help 
distinguish different disks from each other
and determine which physical models are appropriate.

The physical quantities of most interest describe the geometry of the
disk and its microphysics.
We have met $H$, the vertical thickness or scaleheight of the disk, and $r$, the radius of the disk, earlier.
In addition, the rotational angular speed of the disk is denoted by $\Omega$, 
the particles' collision frequency by $\omega_c$, their
velocity dispersion by $c$,
and their size by $a$.  Vertical hydrostatic balance implies that
$c\sim H\Omega$. In a frame moving with the bulk velocity
of the fluid,
the particles' mean free path is hence
$\lambda\sim (\Omega/\omega_c)\,H$.
All of these parameters vary within the disk, in particular with radial location.

Sufficiently dense and cool gaseous disks feature collision
frequencies much greater than the rotation frequency, $\omega_c \gg
\Omega$. It follows that the mean free path is significantly less than
the disk scale height.
In other words, the (vertical) Knudsen number is small: $\mathrm{Kn}= \lambda/H \ll 1$. 
As a result,
equilibrium microphysics is dominated by interparticle collisions;
the phase space distribution of particles approaches a
Maxwellian and the equations of fluid dynamics (or magnetohydrodynamics) are
appropriate \citep{Shu92}. 

These relations are reversed
in galactic disks, debris disks,
 and the tenuous plasma in the very hot regions around some black holes 
\citep{BinTrem08,Reesetal82,Narayanetal98}.
In these settings, $\omega_c \ll \Omega$, and so we have $\mathrm{Kn}\gg 1$.
Collisions between particles are infrequent and the familiar continuum descriptions break down.
In particular, the thermal relaxation timescale is of order, or longer
than, the dynamical timescale, leading to
severe
velocity anisotropies and unusual
momentum and heat transport
\citep[e.g.][]{Toomre64,Brag65,LB67}.
Especially extreme environments are the surpassingly hot
midplanes of AGN and X-ray
binaries, where
the ions and electrons collisionally decouple and possess
temperatures differing by 3 orders of magnitude \citep{Reesetal82}.
The collective behaviour in these systems is 
determined by long-range 
gravitational interactions, in the cases of galactic and debris disks,
and by electromagnetic fields, in the case
of a collisionless plasma,
and
researchers must resort to an appropriate kinetic theory or $N$-body 
simulations.

Often planetary rings fall uncomfortably between these frameworks.
Ring particles undergo some tens of collisions per orbit, and thus
$\omega_c \sim \Omega$ and $\mathrm{Kn}\sim 1$ \citep{ArakiTrem86,WisTrem88}.  
Of course, one can persist with hydrodynamic models for certain problems,
but they cannot capture strong velocity anisotropies nor the
non-Newtonian behaviour of the stress
\citep{GoldTrem78,LatOgi06a}. One must instead
call upon an appropriate dense-gas kinetic theory or $N$-body simulations (see
Chapters by Stewart and Salo), neither of which is straightforward to implement nor interpret.

\begin{table*}
\begin{tabular}{||c || c | c| c| c| c |}
 \hline
 Disk system & $r$ & $H/r$ & $T$  or $c$  & $\omega_c
 / \Omega$& $n$ 
\\
 \hline
 Protoplanetary disks & $>100$ AU & $0.05$ & $10-3000$ K & $10^{7}-10^{12}$ & $10^9-10^{18}$ cm$^{-3}$    \\
 Dwarf novae   & 0.005 AU & 0.01 & $10^3 -10^5$ K & $10^{10}-10^{12}$  & $10^{17}-10^{20}$ cm$^{-3}$   \\
 AGN &  $>100$ AU & 0.001 & $10^5 - 10^{12}$ K & $10^{-6}-10^{10}$  & $10^{10}-10^{17}$ cm$^{-3}$  \\
 Debris disks: dust & $10-1000$ AU & 0.01-0.1  & $10^4-10^5$ cm/s & $10^{-5}-10^{-2}$ & $10^{-12}-10^{-9}$ cm$^{-3}$ \\
 Galactic disks: stars   & $10-100$ kpc  & $0.01-0.1$ & $10^6$ cm/s & $10^{-8}$ & $0.1-100$ pc$^{-3}$ \\
 Galactic disks: gas & $10-100$ kpc  & $0.001-0.01$ & $10-10^4$ K & $10^8$ & $10^{-1}-10^6$ cm$^{-3}$  \\
 Saturn's dense rings & 0.001 AU & $10^{-7}$ & 0.1 cm/s & $ 10$ &
 $10^{-6}$ cm$^{-3}$ \\
 \hline
\end{tabular}
\caption{Characteristic scales and parameters of selected disk
  categories. Estimates are lifted from references given in Section
  \ref{s:survey}. Here $T$, $c$, $\omega_c$, and $n$ refer to
  temperature, velocity dispersion (or sound speed), collision
  frequency, and number density respectively. 
  Ranges of protoplanetary disk properties are for radii
  between 0.1 AU and 100 AU. The AGN estimates include the main disk
  but not the broad-line region at radii $\gtrsim 1000$ AU. The
  extremely hot temperatures (and consequent low collision
  frequencies) correspond to the radiatively inefficient inner disk
  around a $10^7$ solar mass black hole.
  The galactic estimates are for the Milky Way, and do not
  include the extremely hot diffuse phase of the ISM.}
\end{table*}

Dense rings possess an additional peculiarity: not only is the mean free
path of order the vertical size of the system, so is the particle radius, $a$. Thus we have
$$\lambda\sim a\sim H,$$ 
or, formulated another way, $R= a/H \sim 1$, where $R$ is the
Savage--Jeffrey parameter of granular flow theory \citep{SavJef81}.
In short, the particles that constitute the disk 
are macroscopic bodies, another impediment
to a continuum description.
There is no astrophysical analogue for this situation, as 
even the largest bodies in debris disks, 
and certainly stars in galaxies, possess $R \ll 1$ 
\citep{Wyatt08,BinneyMerri98}. 

There are important dynamical consequences when
$\lambda\sim a \sim H$. Excluded-volume
effects impose not only an unusual equation of state, but 
also precipitate the vertical `splashing' of particles out of the ring plane.
In addition, the ring's rheology undergoes a dramatic alteration:
the \emph{collisional} transport of momentum (from one
particle to another during a collision) dominates over the
standard \emph{free streaming} or \emph{translational} transport 
(by individual particles between collisions). The dependence of the
latter on the system parameters and state variables is markedly
different and leads to alternative instabilities and dynamics 
\citep{ArakiTrem86,WisTrem88}.
 
Next we consider the range of scales upon which dynamical
phenomena can manifest. Gaseous disks almost always exhibit 
$$ r \gg H \gg l,$$ 
where $l$ is the viscous scale, defined for most
dynamical purposes by $l =\sqrt{\nu/\Omega}$, with
$\nu$ the kinematic molecular viscosity. 
Using $\nu\sim \lambda c$, we obtain
the scalings 
$$l\sim \sqrt{\lambda H}\sim (\Omega/\omega_c)^{1/2}\,H.$$ 
Another way to put this is in terms of the
 Reynolds number, 
defined by $\mathrm{Re}= H^2\Omega/\nu\sim H/\lambda \gg 1$.
In most gaseous disks, the gulf separating the disk radius $r$
from the disk thickness $H$ is much
smaller than that between $H$ and the viscous dissipation length $l$.
For instance, in a protoplanetary disk at $1$~AU, $H/r \sim 0.05$, and
$l/H\sim 10^{-5}$. The Reynolds number is hence huge $\sim
10^{10}$ \citep{Armitage11}.

Contrast this with Saturn's dense rings where 
$$ r \gg H \sim l,$$
and energy is dissipated on lengthscales of order the disk thickness $H\sim a$. 
There simply are
no shorter meaningful scales.
Planetary rings
are low Reynolds number flows, $\mathrm{Re}\sim 1$,
even if their viscosity is more than four orders of 
magnitude less than that of a 
protoplanetary disk: $10^2$ cm$^2$/s versus $10^6$
cm$^2$/s \citep{Tiscetal07,Armitage11}.

How does this influence the dynamics? 
Consider the onset of instabilities.
In many cases, the input of energy is on a lengthscale around $H$, and
in gaseous disks, on scales consequently much larger than $l$. 
Because of this separation, instabilities typically 
saturate by initiating a \emph{turbulent cascade} of energy to the distant dissipation scales, 
as this is the most efficient way to rid the system of the excess energy. 
Planetary rings make a striking contrast, because $l$ is not far from
the input scale $H$. 
Though the ring viscosity is comparatively 
tiny, the system is `controlled' by dissipation, and 
instabilities saturate instead by \emph{generating structure}
--- and there is a huge range of scales between $H$ and $r$
available in which to do
so. 
The result may be chaotic and disordered, but it is
categorically different from turbulence.
Note that if there is a sufficient separation between $H$ and $r$,
gaseous disks can in principle exhibit structure on
intermediate scales as well: for instance, modulations riding on
small-scale turbulence \citep{Joetal09}.

Finally, an important distinguishing property of gaseous disks, 
at least, is their ionisation fraction. 
Cold and poorly ionised disks undergo very different dynamics compared
with
partially and fully ionised disks
\citep{BlandPayne82,BalbusHawley1991,BlaesBalbus94,Wardle99}. 
This issue is less relevant in comparisons with dense planetary rings, 
which are composed of mostly uncharged macroscopic particles, and are thus
closer to decidedly hydrodynamic systems, such as the cold
neutral regions (`dead zones') of PP disks \citep{Gammie96}.
However, the low-collisional dust in faint rings when charged does suffer 
dramatic qualitative changes wrought by electromagnetic effects. 
The spoke phenomenon in the B-ring
is perhaps the most famous
example, but periodic structures are also forced by Saturn's magnetic
field, and large-scale electromagnetic instabilities have been
postulated \citep[][Hedman chapter]{GM88,Horanyietal04,Horanyietal09}.

\section{FORMATION}

Most disk formation routes draw on either (a) the collapse of a
cloud of material, (b) the tidal disruption
of a body, due to its close approach to another massive body, or (c)
a physical collision. Similarly, ring
formation scenarios fall into one of these three camps, and historically
have strongly influenced, and been influenced by, the question of
disk formation generally. In this section we briefly review the three
ideas. See the Charnoz chapter for a more in-depth discussion.

The first scenario is particularly relevant for proto-planetary
disks (Section \ref{s:PPdisks}), which form from collapsing dense cores within gravitationally unstable
molecular clouds \citep{McKeeOstriker07}. Because the collapsing material
usually carries non-zero angular momentum, it flattens out
and ultimately forms a disk orbiting the protostar. 
Turbulence in the gas then processes the angular momentum
and permits the remaining mass to fall upon the young star.
Note that magnetic fields and non-ideal MHD probably control
the early collapse phase, and possibly the ensuing
turbulent state \citep[e.g.,][]{Joos12}. 
A similar scheme was originally postulated for disk galaxies
\citep{Eggen1962}, though now it is generally accepted that
they form via a sequence of mergers followed by gas contraction into a disk \citep{SearleZinn1978}.

The early stages of Saturn's formation involved a collapsing spinning blob of
gas in the protosolar nebula, some of which must have found its way
into a
circumplanetary disk. One theory posits that Saturn's rings are the
icy leftovers of this disk, the gas long accreted or blown away
\citep{Pollack1976}. In this scenario the rings are primordial, almost
as old as the Solar System, and share a similar dynamical origin (on
a much smaller scale) to their host protoplanetary disk. As explained in
the Charnoz chapter, however, the theory
has problems explaining their icy composition,
in addition to the relative absence of darker material borne
by impacting meteroids (an indication of relative youth)
\citep[e.g.][]{EstradaCuzzi1996}. Perhaps most problematic is the
question of angular momentum, which is being drained from the system
by ballistic transport (amongst other processes), and which
would ensure a lifetime shorter
than that of the Solar System
\citep{Durisen84}.

The second class of disk formation scenarios appeals to the full or partial tidal disruption of a secondary object. 
As described by Roche in the 19th century, a body that orbits too
close to a central mass will be pulled apart by the combined action of
gravitational and centrifugal forces \citep{Chand69}. Examples of
astrophysical disks that form from the disrupted material include
those around supermassive black holes after a stellar disruption, and
the debris surrounding polluted white dwarfs (described in Section
\ref{s:TDEs}). Related, less dramatic, examples are those associated with close binaries, where
the secondary star overflows its Roche lobe, and the resulting stream
feeds an accretion disk around the primary (Section \ref{s:WDs}).
 
Roche's theory, in fact, was originally applied to the formation of Saturn's
rings; he proposed that a moon had veered too close to Saturn and had
been tidally disrupted. 
Recent work augmenting this theory includes that of \cite{Harris84} and \cite{Canup2010}. 
The latter, in particular, posits the tidal disruption of
the icy shell of a
differentiated Titan-sized body and the subsequent
accretion of its rocky core; the theory then naturally accommodates
the striking homogeneity of the rings. 
An explanation similar in some details has been proposed
by \cite{Leinhardt2012} for the dense Uranian ringlets. 
The inward migration of the moons may be driven
by interactions with the circumplanetary disk or tidal
interactions with the planet. A recent passing comet may also have been disrupted
in a similar fashion \citep{Dones91}, but estimates of the cometary
flux at Saturn suggest that this would be an exceptionally rare
occurrence in the last billion years \citep{Lissauer88}.

The third formation scenario involves collisions. 
The astrophysical objects most closely associated with this scenario are
dusty disks around stars (Section \ref{s:Debris}). 
Because of collisional destruction and radiation pressure, this dust
must be constantly replenished by continual but
infrequent collisions between the larger bodies in the belt. 
Collision-based formation theories for Saturn's dense rings
have appealed to a \emph{single} cataclysmic impact between two moons or between a 
moon and a comet, the latter event occurring during the late heavy
bombardment \citep{Pollack75,Charnoz09}. Both scenarios inherit the problems of
the homogeneous composition and apparent youthfulness of the rings. 
It is plausible, however, that the F-ring is the result of a recent
impact between Prometheus and Pandora \citep{Hyodo2015}, and a similar event
may be responsible for the Neptunian arcs \citep{Smith89}.

A more direct analogy with debris disks is provided by the dusty rings in the Solar System. 
These include Saturn's F-ring, whose dust is continually supplied by the
collisions between embedded larger bodies \citep{CuzziBurns88,BarbEspo02}. 
They also include Saturn's G-ring, Methone, Pallene, and Anthe rings,
in addition to the
Jovian dust rings. 
For the most part, the dust in the latter disks issues from
collisions
between extrinsic impactors (micrometeroids) 
and embedded large bodies, such as atmosphere-free moons 
\citep{Hedman2009,Hedman10,Burns99}. 

Finally, we mention Saturn's E-ring and
Io's plasma torus, both vulcanic in origin, for which
there are no perfect extrasolar
analogues. Very recent observations, however, show short-lived arcs
of material around close-in exoplanets, ejected from their
atmospheres or surfaces \citep{Shout14,Sanchis15}.

\section{ACCRETION}

Once an astrophysical disk has formed, its subsequent evolution and ultimate dissipation are,
to a large extent, dominated by accretion. Disks have a finite lifetime and ultimately fall 
onto their central object via a process of angular momentum redistribution. 
The exceptions are galactic and debris disks.
In this section we focus on accretion driven by local angular momentum
transport, 
such as turbulence or interparticle collisions. Note that other
mechanisms exist that are important in certain circumstances, such as magnetocentrifugal winds, 
or large-scale waves \citep{Pudritz07,BulbyPap99}.

\subsection{Gaseous disks}

The classical theory of accretion disks
\citep[e.g.][]{LBPringle74,Pringle81}
describes the evolution of a (more or less)
continuous disk or ring of matter in circular orbital motion around a
massive central body.  Any dissipative process that acts similarly to
a frictional or viscous force causes the inner parts of the disk,
which rotate more rapidly, to transfer angular momentum to the outer
parts.  The disk spreads as its inner and outer parts move to smaller
and larger orbital radii consistent with their specific angular
momentum. Alongside this outward transport of angular momentum there is
a net inward transport of mass. Meanwhile the orbital energy of the disk is
lowered and heat is generated which, once radiated and intercepted
by astronomical instruments, yields some of the observations described
in Section~\ref{s:survey}. Ultimately the disk will fall upon the central
object in a time of roughly $\sim r^2/\nu$, where $r$ is the disk's 
outer radius. According to this estimate Saturn's rings will be gone
in $10^{10}$ years, longer than the age of the Solar System. 
In contrast, if the only momentum transport process in PP disks
was molecular viscosity, they would survive for some $10^{16}$ years,
longer than the age of the Universe! 
To explain the observations in PP
disks at least,
an effective, or `anomalous', viscosity must be present 
at a much greater magnitude, 
some $ 10^{16}$ cm$^2$/s. 

Within the astrophysical disks mentioned in
Section~\ref{s:survey}, several different mechanisms of angular-momentum
transport may be operating.  Some gaseous disks are sufficiently hot
that the main constituents are substantially ionised; these include the
disks around black holes and compact stars in interacting binary
systems (at least during their actively accreting phases), and the
protostellar disks of FU Orionis systems.  In these systems the
magnetorotational instability \citep[MRI;][]{BalbusHawley1991,BalHawl98} is
expected to sustain a dynamically significant magnetic field and to
provide measurable angular-momentum transport through
magnetohydrodynamic turbulence.  In cooler disks, such as typical
protoplanetary systems, where the degree of ionisation is much lower,
the MRI may be restricted only to special regions of the disk \citep[see
for example][]{Gammie96,Armitage11}.  Other, purely hydrodynamic mechanisms have been
proposed to permit sustained activity
in magnetically dead regions.  These include turbulence instigated by gravitational instability 
(`gravitoturbulence'; see Section \ref{i:GI}), 
which may attack the more massive early stages of PP disks \citep{PS91,Durisen07}, 
subcritical baroclinic instability \citep{LP10}, 
vertical shear instability \citep{Nelson13,Barker2015}, and vertical convection
\citep[though it may be difficult to self-sustain;][]{LesOg10}.

In situations where a plausible mechanism of angular-momentum
transport has been identified, the difficulty remains of quantifying
its efficiency.  Most transport processes are stochastic in nature,
but what is needed for the global evolution of the disk is the mean
value of the shear stress and its dependence on relevant quantities
such as the density, pressure, shear rate, etc.  \citet{SS73}
introduced a useful parametrisation in which the shear stress
is written as the pressure multiplied by a dimensionless parameter
$\alpha$.  In the case of hydrodynamic or magnetohydrodynamic
turbulence, $\alpha$ is expected to lie between $0$ and $1$
 if the perturbations of the fluid velocity (and Alfv\'en
velocity, in the MHD case) are related to the sound
speed (but typically less than it).
This `alpha-disk'
prescription has dominated accretion-disk theory for some 40 years, as it
permits researchers to construct models of disk evolution and structure,
and hence generate synthetic emission spectra that can be
compared with observations.

\begin{figure}[ht]
\figurebox{15pc}{}{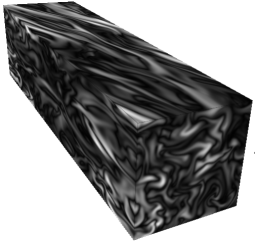}
\caption{A snapshot from a simulation of MRI turbulence conducted in
  the shearing box model of a gaseous accretion disk. The
  field plotted is the magnetic field strength $|\mathbf{B}|$. Credit: Tobias
  Heinemann.
\label{fig:MRI}}
\end{figure}

The turbulent state can be investigated through numerical simulations
of a relevant system of hydrodynamic or magnetohydrodynamic equations.
If the disk is thin and the correlation length of the turbulence is
small compared to the orbital radius, then a local simulation based on
the shearing box \citep[e.g.][]{HawlBalb95} is usually sufficient;
the shear stresses can be
measured directly from the simulation. Figure \ref{fig:MRI} presents a
snapshot from a local simulation of the MRI. Attempts have also been made
to describe the turbulent state analytically by means of a set of
moment equations derived from the basic hydrodynamic or
magnetohydrodynamic equations using a closure model, such as a simple
modelling of the triple correlations of velocity and magnetic
fluctuations \citep{KY93,KY95,Ogilvie03}.

\subsection{Rings}

In the case of planetary rings, angular momentum is transported in
part by a viscous stress associated with an anisotropic velocity
distribution of the particles, and therefore with an anisotropic
pressure tensor (free-streaming transport).  
Indeed, the $\alpha$ viscosity parameter of a dilute
planetary ring (and also a dilute plasma) can be linked to the degree
 of anisotropy in the
pressure tensor. 
 In dense rings, however, there is an additional
contribution from the transport of angular momentum during
(rather than between) collisions. 

 There is a close analogy between
the behaviour of the pressure tensor in a dilute planetary ring and
the Reynolds stress tensor that describes the correlations of velocity
fluctuations in a turbulent gaseous disk \citep{QC00}. 
These tensors obey evolutionary equations with some identical terms,
describing the interaction of the fluctuations with the Keplerian
orbital motion (which tend to make the tensor anisotropic),
as well as some differing terms, describing the collisional and nonlinear
dynamics.  In the case of a dilute planetary ring, collisions
make the pressure tensor isotropic and damp the fluctuations.
\citet{GoldTrem78} showed that a circular dilute equilibrium can be
sustained if the collisions are sufficiently elastic.  Similarly, in
purely hydrodynamic turbulence, the nonlinear effects tend to both isotropise
and damp the velocity fluctuations.  Sustained hydrodynamic turbulence
is possible only if the isotropisation is relatively strong compared
to the damping, a condition that is not believed to be satisfied in a
Keplerian disk \citep{Ogilvie03}. Note that additional thermal gradients may permit
sustained activity.

An additional complication is that planetary rings, like gaseous disks, can suffer instabilities that lead to 
disordered flows on large scales ($>H$), 
e.g.\ gravitational instability \citep{Salo92,Salo95}. 
The correlated motions of the associated turbulence transport 
angular momentum up to an order of magnitude greater than both the free-streaming 
and collisional stresses \citep{Daisaka01}.
This transport can also be parametrised by an alpha prescription, though such a model must omit the complicated interplay between 
the turbulent wakes and the collisional dynamics of the ring particles, the length 
and timescales of which are not well separated. Plasma systems that undergo analogous mixed behaviour 
involving `micro-turbulence' are the solar wind,
the intracluster medium, and the inner regions of black-hole accretion disks \citep[e.g.][]{Kunz14}.
 
The alpha model, and more sophisticated approaches, are mean-field
theories in which the details of the small-scale physics (turbulent motions, collisions) are 
deemed quasi-steady and averaged away.
However, for certain astrophysical processes 
this microphysics cannot be treated so indifferently.
Planet formation and dust production, in protoplanetary and debris
disks respectively, 
rely on the details of collisional disruption 
and agglomeration, which are also relevant for both dense and dusty planetary rings. 
In the next section we review these processes.

\section{COLLISIONS AND DUST}

The collisional dynamics of Saturn's and Uranus's macroscopic ring
particles are unusual in astrophysics on account of
their low impact speeds $\sim$ mm/s.
In fact, these
are of order the escape speeds from the larger
particles, and only enhanced tidal forces (due to their orbit)
prevent gravitational collapse.
Solid bodies
in debris or protoplanetary disks collide more violently on the whole,
at speeds ranging between
mm/s and km/s.
Nevertheless, there is significant overlapping physics that is
instructive to review.
We also discuss in this section the connections between the 
dynamics of dust in debris disks and the tenuous rings of the Solar System.

\subsection{Planet formation}

The theory of planet formation tracks the agglomeration of
solid particles in disks from micron-sized dust to the $10^3$ km cores of giant
planets. Vaulting this tremendous gulf, spanning some 12 orders of magnitude,
requires multiple growth mechanisms and physical processes.

Microgravity experiments show that impacts
between sub-centimetre dust grains are controlled by
their coupling with the ambient gas and generally 
result in sticking collisions, whether their motions are
Brownian or induced by turbulence \citep{BlumWurm2008}. 
Typical impact speeds are of order 1 
$\mathrm{m}$/s \citep{Brauer08} and attractive surface forces are relatively
strong in this regime.
Larger
particles are only weakly coupled to the surrounding gas flow and 
thus achieve greater impact speeds, meaning
surface forces become less dominant.
As a consequence,
collisions are more likely to result in bouncing or fragmentation, not sticking, and
agglomeration to sizes larger than centimetres is difficult.
Note that these details
are complicated by particles' structure
(how `fluffy', how compacted, etc.) 
and their composition (silicate versus icy, for instance) \citep{Testi2014}. 
Especially interesting are collisional outcomes between
solids of disparate sizes, which sometimes result in a net mass transfer
to the larger particle. Hence there may exist a narrow route by which a small
number of `lucky' aggregates can grow to km sizes, leaving
behind a swarm of small `unlucky' grains \citep{Wurm2005,Testi2014}. 

\begin{figure}[ht]
\figurebox{20pc}{}{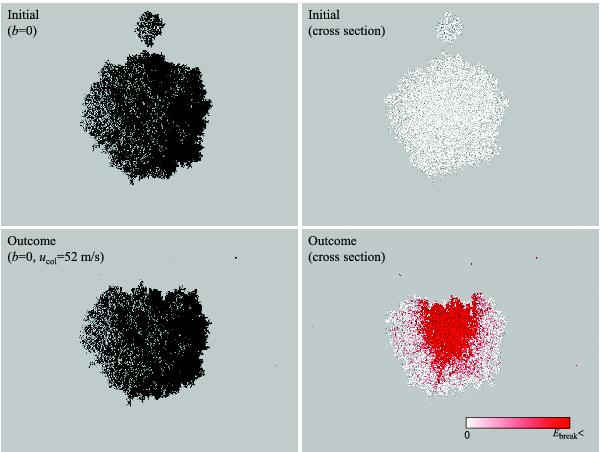}
\caption{Snapshots before and after a simulated collision between dust
  aggregates. The red shading in the right panel indicates the amount of
  energy dissipated during the impact. The number of spherical
  components in each cluster is 2000 and 128,000 respectively. 
 The impact speed is
  52 m/s. Credit: \cite{Wada2013}. \label{fig:Dust}}
\end{figure}

In conjunction with laboratory experiments, 
numerical simulations, using $N$-body molecular dynamics and SPH, 
have
determined collisional outcomes between aggregates of up to decimetre 
sizes \citep{Wada2008,Geret2010,Ringl12,Seiz13}. Figure
\ref{fig:Dust} shows the outcome of such a simulation. Additionally,
statistical information has been gleaned from integration of
suitable coagulation equations \citep[e.g.][]{Safronov69,Dohn69,Tanaka96}
that evolve forward in time the
distribution function of a swarm
of colliding dust grains
\citep[e.g.][]{DullDom2005,Windmark12,Garaud13}. In such calculations,
collisions are parametrised in a mathematically convenient but
also physically motivated way. 

\begin{figure*}
\figurebox{40pc}{}{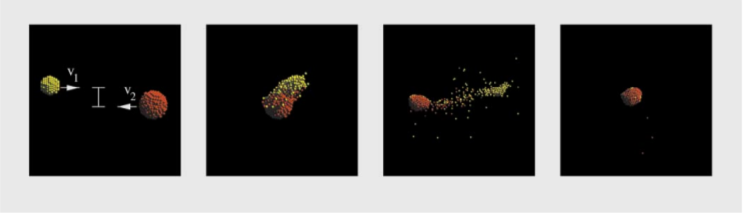}
\caption{Snapshots before, during, and after a collision between two
  planetesimals using an $N$-body simulation. Credit: \cite{Zoe2002}. \label{fig:planets}} 
\vspace{3pc}
\end{figure*}

The growth of larger bodies, from centimetre to kilometre sizes,
 is an especially active area of research. Most
theories rely at some point on the gravitational 
collapse of many particles, usually in
collaboration with aerodynamical effects: streaming instability;
accumulation in `dust traps', such as zonal flows and disk vortices;
`pebble accretion', etc. 
\citep{Youdin2005,Barge95,Lambrechts12}. The robustness and efficiency
of these various mechanisms are still unclear.

Solid bodies
above roughly a kilometre (called planetesimals)
are decoupled from the gas and
further coagulation is achieved by direct collisions, the frequency of
which is aided significantly by gravitational focusing
\citep{PapTerq06}. Numerical SPH and $N$-body simulations
have probed the outcomes of their collisions and show that they
 fall
into a variety of regimes: cratering, merging, fragmentation,
`hit-and-run', and
annihilation \citep{Leinhardt12b}.
 These regimes partly depend on whether
the bodies are held together by tensile strength or self-gravity. 
 Figure \ref{fig:planets}
shows snapshots of a simulation of a high-velocity collision between
two planetesimals.

The collective dynamics of a swarm of planetesimals may be modelled
with a suitable coagulation equation or $N$-body simulation
 \citep[see e.g.][]{WetherillStewart93,KokuboIda96,
Weidenschilling97,Richardson2000}. These typically indicate runaway
growth of a few aggregates which halts upon reaching 1000 km 
sizes \citep{Greenberg78,WetherillStewart89}.
The resulting planetary `embryos' or `protoplanets' continue to acrete, albeit
at a much reduced rate, via what
is termed oligarchic growth \citep{KokuboIda98}.

\subsection{Debris disks}

If researchers in planet formation focus almost 
exclusively on how large objects
are assembled from dust grains, one could say researchers of debris disks
take the diametrically opposed viewpoint: how do large bodies produce
the observed tiny grains? In fact, the process of runaway accretion in
planetesimal belts produces not only larger objects, but also significant
quantities of dust \citep{KenyonBromley2004b,KenyonBromley2004a}.
Dust production hence continues well after the disk gas
dissipates and throughout the intermittent collisional evolution of the 
`leftover' planetesimals and protoplanets \citep{Wyatt08,Matthews14}. 

In contrast to the coagulation equations employed in planet formation,
researchers compute the statistical distribution of debris disk
solids with collisional cascade models. The largest
bodies ($\sim 1-100$ km) are input as `fuel', and mass is lost
at the smallest sizes due to radiation effects and collisional
destruction. 
The resulting wide dynamical range, some 40 orders of magnitude in
mass, makes these calculations especially difficult.
The simplest models assume a steady-state
size distribution but with decreasing total mass
\citep{DD03}. For detailed comparison with observations,
however,
more advanced variants are needed that include, for instance, 
the 
particles' orbital elements, and use kinetic theory \citep{Krivov06,Loehne12} 
or `particle in a box'
methods \citep[e.g.][]{Thebault07}. 

$N$-body codes that track explicitly each member of a
 small subset of the solids have been useful in simulating the effects
 of large perturbations on disk structure, 
such as those arising from an embedded
 planet. They struggle, however, to explain the overall distributions.
Hybrid codes have emerged recently that
comprise $N$-body simulations coupled to dust evolution, thus describing
both dynamics and collisions accurately and concurrently
\citep{Kral13,Nesvold13}. The codes evolve forward in time
the properties of a cloud of similarly sized 
particles on the same orbit (`superparticles'). 
Collisions between these groups 
generate new superparticles that represent the post-collisional
fragments \citep{Matthews14}.

\subsection{Rings}
\label{s:ringscollision}
As in the fields of planet formation and debris disks, the collisional
dynamics of planetary ring particles has been explored with 
laboratory experiments and statistical methods. The overwhelming 
majority of
work, however, has been undertaken with $N$-body simulations. Owing to numerical
limitations, these have focused on 
shorter time-scale dynamical phenomena such as the
onset of instabilities and satellite wakes
\citep{Salo91,Salo01,Lewis2000,Lewis2009}, rather than 
on the slower processes that shape the particles' size distribution.
Most $N$-body studies of dense rings involve hard indestructible
spheres, either identical or with a fixed distribution of sizes. Collisions 
are controlled by a normal coefficient of restitution, and a
tangential coefficient of restitution when including the particle
spin. See Chapters by Stewart and Salo for further details. Needless to say,
the regime of frequent and gentle collisions, which characterises
dense rings, is very remote from the contexts described previously in
this section.

Because of the low impact speeds, proximity to the planet, and the
properties of the particles' regolith, a myriad of processes control the
evolution of the size distribution in dense
rings. In addition to inelastic `bouncing' (modelled in $N$-body
 simulations), collisions may lead to: adhesion (due to the
 meshing of micron-sized surface structures or more drastic structural
 reconfigurations); surface compaction of loose frost
 (`polishing');
 mass transfer between impacting particles; as well as
 the more familiar collisional fracture. Laboratory experiments have
 been essential in uncovering and 
characterising these various effects 
\citep[e.g.][Colwell Chapter]{Bridges84,Hatzes88,Hatzes91,Supulver95,Supulver97}. 
A raft of non-collisional processes also contribute. These
include gravitational recapture of collisionally dislodged fragments,
tidal fragmentation, rotational fragmentation, and erosion by
micrometeoroid impacts \citep{Weidenschilling84}. 
This miscellany of physics includes effects present in protoplanetary
dust dynamics (bouncing, adhesion, polishing, fragmentation) and
planetesimal dynamics (rotational fracture, gravitational recapture),
as well as new effects (tidal fracture, micrometeoroid bombardment). 

In parallel to laboratory experiments, theoretical descriptions of
individual collisions have been developed that employ viscoelastic
theories \citep[e.g.][]{Spahn95,Albers06}.
On the other hand, computing the energetics of variously packed aggregates
can characterise collisional outcomes as a
function of impact speed \citep{Guim}. Future work, involving $N$-body
or molecular dynamics simulations 
\citep[as with planetesimals;][]{Leinhardt12b}, may categorise collisional
events more securely. 

There exist a small number of statistical studies exploring
the cumulative effect of 
collisional coagulation and fragmentation, such as \citet{Weidenschilling84}, 
\citet{Longaretti89}, and more recently 
\citet{Bodrova12} and \citet{Brilliantov15}. 
Finally, `sticky' collisions have been
modelled in a restricted set of $N$-body simulations that
numerically produce
self-consistent size distributions in relatively good agreement
with observations
\citep{PR11,PR12}. 
In comparison with the
field of planet
formation, however, this area of research, 
though very promising, is underdeveloped. For example, no well-defined
`barriers' have been computed above which growth of large aggregates halts,
and below which small particles are efficiently swept up by larger
ones. 
Nor have calculations been attempted that could decide if it is statistically likely
that a few `lucky' aggregates could grow to very large sizes $\sim
100$ m (as in planet formation theories). If such growth was possible,
it may provide an explanation for the observed propellers in Saturn's
A-ring (see Spahn chapter).

The collisional dynamics of Saturn's F-ring differs from that of the inner
dense rings.  
The F-ring consists of a belt of large
$\sim$ 1 km bodies swathed in the dust generated by their mutual
collisions \citep{CuzziBurns88,Colwelletal09,Attree2012,Meinke12}. The larger
bodies are possibly the fragments of a catastrophic collision between
Prometheus and Pandora \citep{Hyodo2015} that
 are in the slow process of being ground
down collisionally \citep{CuzziBurns88}, a scenario directly analogous
to models of debris disks. 
Neptune's dusty rings, which are far less well studied, probably 
share a similar formation history and similar dynamics \citep{Smith89}. Another
connection is the powerful influence of a nearby satellite. Prometheus
significantly `stirs' the F-ring material, enhancing the
collision rates of the $\sim$ km sized moonlets and perturbing the
overall structure of the ring in much the same way that embedded or nearby planets
shape the dust in debris disks \citep{Murray05,Beurle10,Attree14}. 

One of the essential features of the F-ring is its weaker tidal
environment, as compared to the inner dense rings. Gravitational
aggregates at smaller radii have difficulty growing to large sizes
before they are tidally disrupted (an interesting contrast to
debris disks and planetesimal belts). 
This has stimulated alternative
theories for F-ring dynamics that posit that the population
distribution is quasi-static, the number of
large bodies regulated by fragmentation and 
gravitational accretion
\citep{BarbEspo02}. Indeed it is true that
 $N$-body simulations show
F-ring particles readily clump into
gravitationally bound aggregates, akin to `rubble piles' 
\citep{KSalo04,LRO12},
whose further growth and collisional destruction
characterise the general dynamics \citep{K07,Hyodo14}. The collective outcome of
aggregation and disruption has been theoretically explored
 using statistical methods similar to those employed
in other astrophysical
disks \citep{BarbEspo02,Esposito12}. There remains, however, plenty
of scope to further 
apply the well developed techniques of debris disks and planet formation to this
problem.

\subsection{Radiation and other forces}

In debris disks, radiation forces \citep{Burns1979}
 play an important role in the dynamics of small particles.
Often a distinction is made between radiation pressure and
Poynting--Robertson (PR) drag, although the origin of the two effects
is the same (photons
 transferring energy and momentum to dust grains).
Radiation pressure removes micron-sized dust grains
from the system on a dynamical timescale, thus enforcing a strict
lower cut-off in the size distribution of solids \citep{Wyatt08}.
PR drag, on the other hand, causes small particles to spiral into the
central star, but on a timescale longer than collisional erosion. It
is hence far less important \citep{Wyatt05a}.

In contrast, the PR drag on planetary ring particles can lead to
rapid inward migration of dust \citep{GT82}, especially an issue in
the F-ring and Saturn's diffuse rings \citep{Sfair09,Verbiscer09}.
This is less important in the inner dense rings where the dust
is partially shielded from the Sun and its dynamics controlled
by collisions with larger particles.
The scarcity of sub-mm dust in dense rings can be attributed to
their efficient absorption by larger particles \citep[Becker et al.\ 2016,][]{French2000}. 

Ring dust is also
subject to drag forces issuing from both the host planet's exosphere and the
dilute plasma coorbiting with its magnetosphere. As a result, dust can feel either
a headwind or tailwind that causes radial migration and
eventual loss from the system, an effect remarkably similar to the
aerodynamical migration experienced by centimetre to metre-sized particles in
protoplanetary disks \citep{Weidenschilling77}. Uranus's outer
atmosphere is especially extended \citep{Broadfoot86}, explaining the paucity
of dust in its ring system, and which may also have some dynamical consequences 
on ringlet confinement \citep{GP87}. Charged dust elevated above Saturn's main
rings, on the other hand, interacts with the Saturnian magnetosphere
leading to radial drifts, angular momentum exchange, potential
instability, and the striking spoke phenomenon 
\citep{GM88,Horanyietal04,Horanyietal09}.

\section{WAVES}

The study of waves (and instabilities) in astrophysical disks began with
attempts to explain the appearance of spiral structure in galactic
disks \citep{Toomre64,LinShu64,GL65}.  
The observed spirals may be interpreted as `density waves', 
a collective phenomenon combining inertial (epicyclic)
forces and self-gravity, but strongly influenced by the orbital
shear.  
Density waves have also been studied in gaseous disks, where they can
be thought of as inertial--acoustic waves because pressure usually
dominates self-gravity, leading to qualitative
differences in their propagation (the group velocity differs in sign,
for instance).

\begin{figure}
\figurebox{22pc}{}{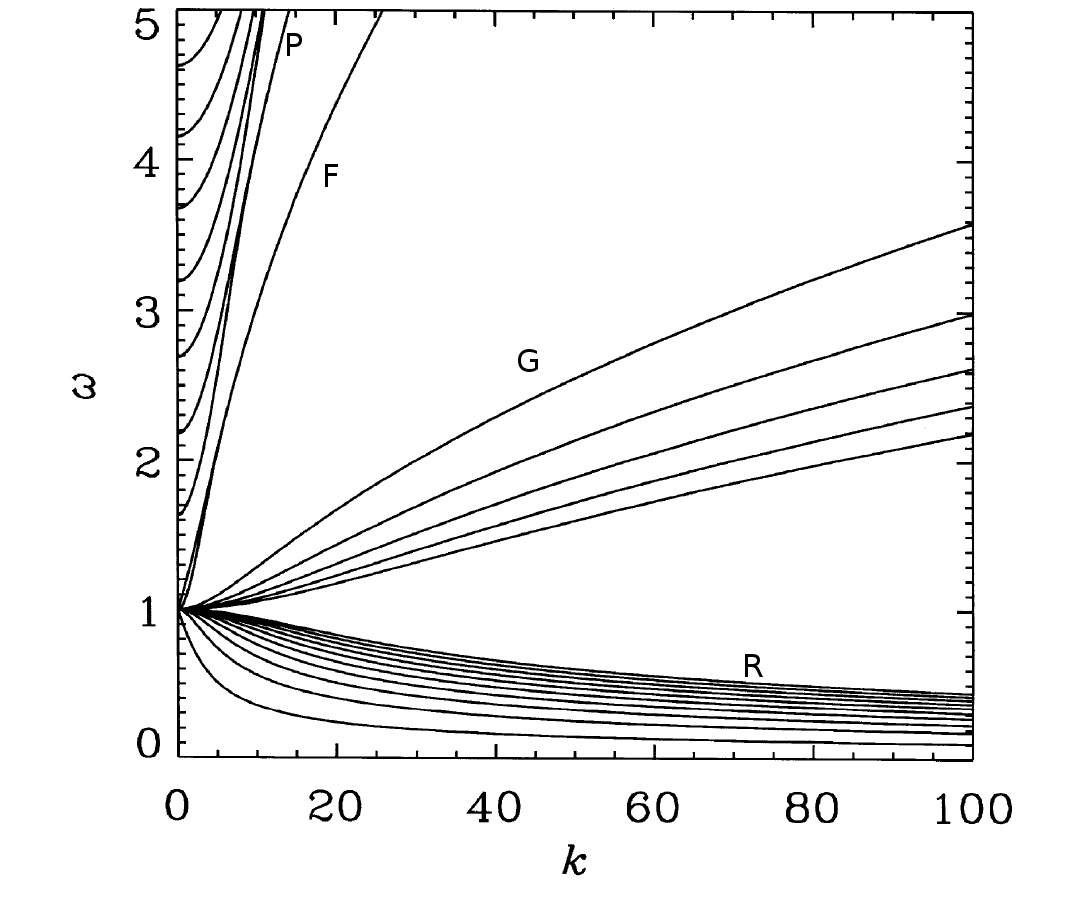}
\caption{The dispersion relations of the first few axisymmetric p, f, g, and r
  modes in a local model of a disk. The frequency
  of the modes $\omega$, scaled
  by $\Omega$, is plotted versus radial wavenumber $k$. The disk is a
  stably stratified polytrope. Credit: \citet{Ogilvie98}.} \label{modes}
\end{figure}

If the viscosity parameter
$\alpha\ll1$, then gaseous disks manifest a variety of additional
wave modes with wavelengths
comparable to or less than $H$. 
Such waves are only accurately
studied in three-dimensional models that resolve the disk's vertical
structure \citep{Loska86,Oka87,LPring93,KP95,Ogilvie98}.
The disk may then be understood as a wave guide through which
the various modes propagate. In some models the hydrodynamic waves can be classified
into f (fundamental), p (pressure) and g (gravity) modes by analogy
with stellar oscillations \citep{Ogilvie98}. Rotation also introduces
low frequency r modes (also called `inertial waves'). Example dispersion relations of these
modes are plotted in Figure \ref{modes}.
Global oscillations can be formed when these waves reflect from radial 
boundaries of the disk or internal turning points.  The study of this rich assortment
of modes is sometimes called `diskoseismology'. One of its aims
has been to explain the curious quasiperiodic oscillations
(QPOs) exhibited by certain X-ray sources \citep{RemillardMcClintock2006},
then use these to probe the
relativistic gravitational fields associated with black holes and neutron stars
\citep[e.g.][]{Wagoner99}. 

The symmetric f modes exhibit the
least vertical structure and correspond to
the spiral density waves observed in galaxies and PP
disks, in addition to the
large-scale eccentric modes inferred in both close
binaries and
certain PP disks. 
In addition, antisymmetric f modes
can manifest
as bending
waves, which transmit a warp (or vertical deformation) through the
disk \citep{PapLin95,Ogilvie99,Ogilvie06}. Vertical tilts and warps,
in fact,
have been observed in X-ray binaries, AGN, and protoplanetary disks,
and are usually driven by a misaligned companion 
\citep{Katz73,Kotze12,Miyoshi95,Marino15}.

In planetary rings the smallest meaningful
lengthscale is the particle size $\sim H$ and hence 
wave modes express little to no
vertical structure: the disk is effectively two-dimensional.
As a consequence, planetary rings cannot support
nearly the same variety of waves as gaseous disks, and
in fact only the f modes
are present.
The \textit{Voyager} and \textit{Cassini} 
spacecraft imaged many examples of spiral
density and bending waves excited in Saturn's A-ring and B-ring by the planet's
moons \citep{Col07}. Their physics is very similar to the
case of close binaries and embedded planets in PP disks 
(see Section \ref{SDinteractions}).

In contrast, some density waves located in the C-ring
may have been generated by low-order normal-mode oscillations within
Saturn itself \citep{Marley91,HedNic13,Fuller14}.
The study of such waves (`kronoseismology') may provide clues
about the internal structure of the central planet. This effort provides a
nice parallel to
diskoseismology's attempts to characterise black
holes and neutron stars.

Large-scale `corrugation
waves' with wavelengths
$>30$ km have also been
observed in the C and D-rings generated by a cometary impact
dating from the 1980s \citep{Hed07,Hed11}. Similar waves were
excited in Jupiter's main ring by Shoemaker--Levy 9 in 1994 \citep{Show11}.
In fact, an analogue of this process occurs in young
protostellar disks: infalling material from the star's natal envelope
can cause spiral shocks in the disk, which may transport a non-trivial
amount of angular momentum, while
thermally processing chondrule precursors
\citep{BossGraham93,LesHenn15}. An important distinction, however, is that the
collective forces are far more significant in gaseous disks, so that
the disturbances are bonafide waves, 
unlike in the ring context, where the structures are essentially
kinematic.

The observed spiral waves in planetary rings possess radial wavelengths that are much
greater than the thickness of the rings $>10$ km, and this means that
self-gravity dominates pressure
(or velocity dispersion) in their propagation, while
the viscosity of the rings usually damps the waves. But density waves have
been observed on much
shorter scales as well. Both axisymmetric and non-axisymmetric density structures
appear with roughly 100 m wavelengths in RSS, UVIS, and VIMS observations
\citep{Thom07,Col07,Hed14}. The
small-scale non-axisymmetric wakes, in particular, give rise to a 
striking large-scale effect, the azimuthal brightness asymmetry 
\citep{Camichel58,Colombo76,Thomp81,Salo92}.

In order to reach dynamically important (and observable)
amplitudes, waves must grow to nonlinear
strengths.
As mentioned, periodic forcing due to an orbiting companion naturally
excites
waves and other, non-propagating, disturbances. 
This is reviewed in Section \ref{SDinteractions}.
Conversely, numerous
instabilities can drive wave (and other) activity to observable
levels. Relevant instabilities are discussed in the following section.

\section{INSTABILITIES}

Because of their physical complexity, gaseous accretion
disks accommodate a large number of instabilities, usually drawing
their energy from the background orbital shear but sometimes also
from vertical shear,
thermodynamic gradients, or directly from the self-gravitational
potential of the disk. It might be said there are more
 instabilities than observations they could feasibly
explain! This is in contrast to Saturn's dense rings where there
is an abundance of observed structure, much of it presumably generated by
instabilities not yet identified or understood. In the following subsection we review only
those processes that are shared by, or have something in common
with, those appearing in planetary rings. 

\subsection{Gravitational instability and gravitoturbulence}
\label{i:GI}
As explained in the Stewart chapter, 
self-gravity decreases the squared
frequency of density waves, leading to axisymmetric instability on
intermediate wavelengths when Toomre's parameter $Q = \kappa c / \pi G
\Sigma < 1$ \citep{Toomre64}. Here $\kappa$ is the disk's epicyclic frequency and
$\Sigma$ is its surface density. 
Note that the criterion given here is for a two-dimensional gaseous disk, and differs slightly in other models.
The above condition can plausibly be met in spiral galaxies,
more massive PP disks, accretion disks in
active galactic nuclei, and, of course, dense planetary
rings.

Gravitational instability (GI) is thought to power the
density waves observed in flocculent spiral galaxies (and possibly
grand design spirals), and thus
controls a crucial aspect of their structure.
Observed spirals in protostellar disks may share the same origin, 
though embedded planets could also drive
these features. GI also features in the `disk instability' theory
of planet formation, by which gas giants are formed by direct collapse
of the disk \citep{Kuiper51,Cameron78,Boss98}. 
In planetary rings, GI is responsible for
`wake' activity on much smaller relative scales, on account
of the extreme thinness of the rings. Because
unstable waves
emerge on scales $\gtrsim H$, they are usually global features in
gaseous disks and local features in planetary rings.

\begin{figure}
\figurebox{19pc}{}{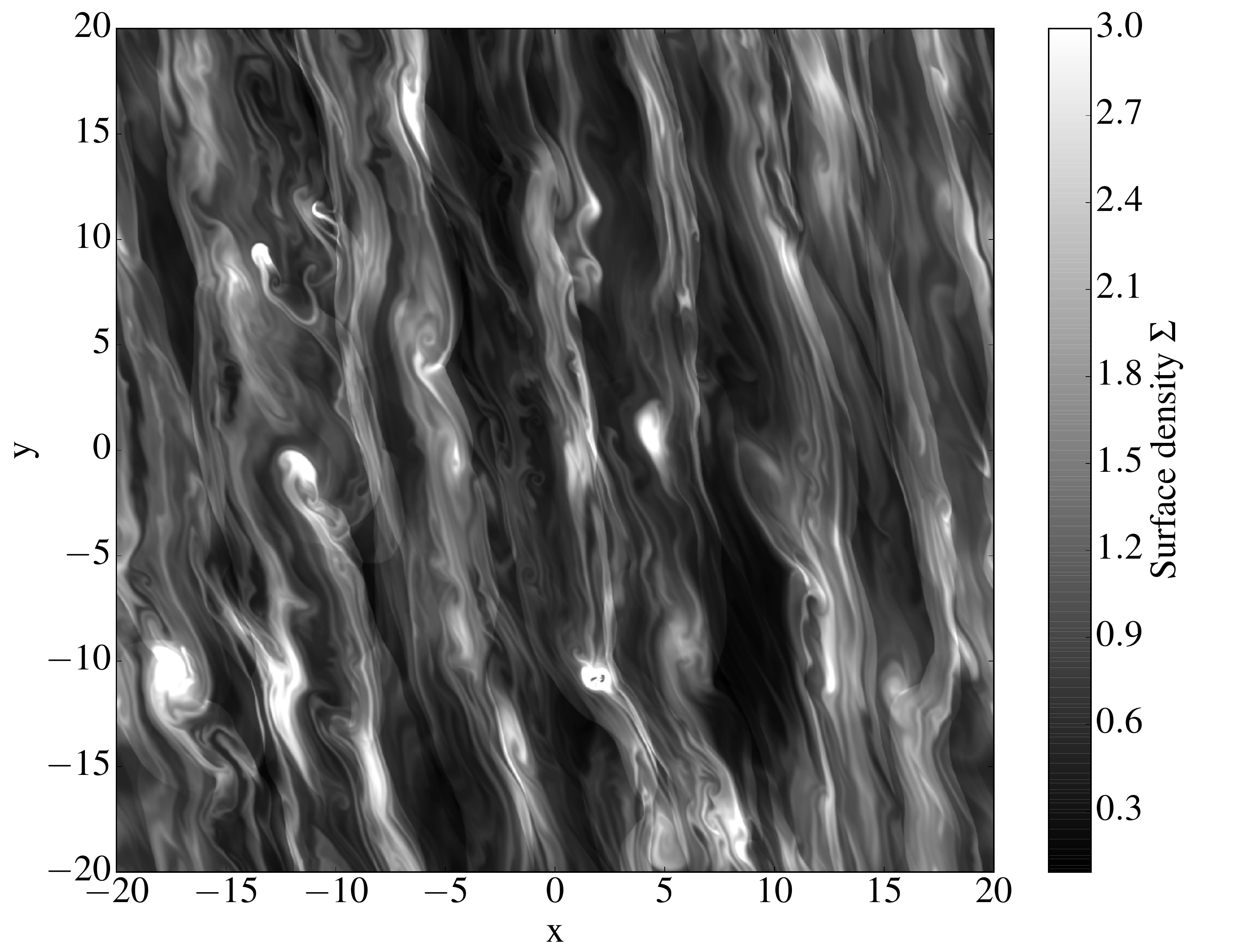}
\caption{Gravitoturbulence in a 2D shearing box model of a gaseous
  disk. The fractional surface density perturbation is plotted, while
  the unit of length is the initial unperturbed scale height
  $c/\Omega$, 
  which here is five times the
  Jeans length $G\Sigma_0/\Omega^2$, where $\Sigma_0$ is the mean
  surface density. The (linear) cooling time is $9/\Omega$. 
Credit: Antoine Riols-Fonclare.\label{fig:GI1}}
\end{figure}

Unstructured disks are linearly stable for $Q>1$, as they cannot
support non-axisymmetric GI modes.
However, for  somewhat
larger $Q\approx 2$, finite-amplitude perturbations instigate
sustained spiral density waves and the system settles into a
`gravitoturbulent' state. Figure \ref{fig:GI1} shows a snapshot of
GI-induced turbulence in a local model of a gaseous disk; here the
mean $Q$ is 2.5.
At least locally, this is a `subcritical'
transition: the disk can support both a laminar
 and a turbulent state for $1<Q\lesssim2$ but leaves the laminar 
equilibrium when given a sufficiently vigorous perturbation.
In practice, the critical amplitude
is small; the shot noise inherent in $N$-body simulations is
always sufficient to disrupt the laminar state.

The nonlinear outcome of gravitational
instability is sensitive to heating and cooling because
$Q$ is proportional to the velocity dispersion, or sound speed, of the
disk. 
 Typically the instability leads to enhanced dissipation that
tends to increase $Q$, so a thermostatic regulation can be achieved in
which $Q\sim 1$ and a stochastic field of density waves
maintained. 
Such an equilibrium has been obtained in local numerical
simulations of gaseous disks \citep{Gammie01}, and the resulting
turbulence has properties in common with states found in
simulations of self-gravitating planetary rings 
\citep[e.g.][]{Salo95,Daisaka01}, even though the
cooling processes differ. 
One important issue for
accretion disks has been the
question of turbulent angular-momentum transport by GI. 
Can gravitoturbulence provide an effective alpha, as
required by the Shakura-Sunyaev theory? 
While it certainly can lead to accretion, it is not always assured
that GI dissipates energy locally because
of its wavelike character \citep{BulbyPap99}, though this seems to be only 
an issue for thicker disks (and certainly not for planetary rings). 

If the disk is permitted to cool on a timescale that is sufficiently short
compared to the orbital timescale, then
the disk may fragment and form a number of bound objects
(stars, planets, moonlets, etc). 
In recent years, much effort has gone
into determining the critical cooling time from
hydrodynamical simulations \citep[e.g.][]{JohnGammie03,Rice05,Durisen07}.   
The question is still open, as it is increasingly clear that
the problem is sensitive to the numerical details of the 
calculations \citep{Paard12,Rice14}. 
Clumps also form in local simulations of planetary
rings \citep[e.g.][]{KSalo04}, 
especially at larger radii where the tidal forces are weaker. 
Generally, however, stable clumps are
more difficult to form in rings, in part because the minimum
lengthscale $a$ is $\sim H$, and clumps are thus
easier to be ripped apart
 by tidal forces and/or collisions with other
particles (or wakes) than the much smaller gaseous cores.
Another way to think about this is in
terms of the equation of state, or a change of phase. Gravitational collapse, in any
system, should end once material becomes so dense that
pressure resists infall
and/or its cooling radically diminishes (via an opacity jump, for
example). In gaseous disks the 
lengthscale upon which collapse halts is exceptionally small $\ll H$, whereas
in planetary rings it is $\sim H$. Once ring material clumps
it almost immediately
becomes `incompressible': it changes state from a `granular gas' to a
`granular liquid'. In addition, cooling is minimised because within a
crowded
 aggregate `collisions' are very gentle (if not
absent) and hence more elastic.
Nonetheless, clump formation is an important feature in the outer
A-ring and in the F-ring, where there is indeed a population
of larger objects, `propellers' and `kittens' respectively
(see Section \ref{s:ringscollision} and chapters by Spahn and Murray).

\subsection{Viscous overstability}

A viscous stress need not just damp density waves (or f modes), it can
also, somewhat counterintuitively, cause such waves to grow. A density wave
 produces stress perturbations that couple the background orbital
shear to the wave velocities. Energy is extracted and the wave
amplified if the velocity and stress oscillations are sufficiently
in phase. The instability typically 
saturates in a quasi-steady state, 
with the viscous driving of the
waves balanced by their viscous destruction (see Stewart chapter). 
The process was first discovered in the accretion-disk
context \citep{Kato78,Blum84}, where it was
hoped it might explain observed luminosity fluctuations around
compact objects. 

As with density waves excited by GI, viscously
overstable modes, and their saturation, 
are essentially global in the accretion disk context.
Thus the inner and outer boundaries, Lindblad resonances, and the disk's
vertical structure all feature in the evolution of the
instability \citep{PapStan86,Kley93,Miranda15}.
In contrast, viscous overstability in planetary rings can be a very
local phenomenon --- certainly when axisymmetric. The fastest growing
modes favour scales of some 100 m, a tiny fraction of $r$ 
\citep{STsch95,Stsch99,Salo01,Schmidt01}, and 
as a
consequence, the saturation mechanism is controlled by nonlinear 
travelling waves \citep{SS03,LO09,LO10,RL13}. 
It is generally accepted that the fine-scale axisymmetric
density waves in Saturn's A and B-rings are generated and sustained by
viscous overstability \citep{Thom07,Col07,Hed14}. 

Viscous overstability can generate large-scale features, such as
eccentric modes \citep{Bord85,Pap88,Long95}. 
It is likely that the structure of eccentric
ringlets and the outer B-ring edge are partly sculpted by this
process \citep{Spitale10,Nicholson14}. 
Similarly, certain gaseous accretion disks may have obtained their
eccentricity in this way \citep{Lyu94,Ogilvie01,LO06}, 
though the issue is far less clear cut than in the
planetary ring context.

One reason why the onset of viscous overstability is problematic in
gaseous accretion disks, as opposed to planetary rings, concerns the
nature of the viscous stresses. In gaseous disks these are presumed to be supplied by
hydrodynamic (or magnetohydrodynamic) turbulence --- but it is unclear if 
the turbulence responds in the desired way when a density wave
propagates through the disk. The stress may not be enhanced in high
density-wave crests, and even 
if it is enhanced the stress may be out of phase with the dynamical
oscillation \citep{Ogilvie03}. 
In either case, viscous overstability may fail to occur. 
Note that the collisional stress in dense planetary rings
does not suffer from this shortcoming \citep{ArakiTrem86,Salo01,LO08}. 

\subsection{Viscous instability}

The viscous instability (also called the `inflow instability') 
occurs when the stresses in a disk decrease
with increasing surface density. The equation for small perturbations in the surface density becomes, as
a result, a diffusion equation with a negative diffusivity. 
Physically, a localised bump of density accretes
less vigorously than its surroundings and mass piles up at its
outer edge, enhancing the overdensity.

In the early 1970s the viscous instability was shown to afflict
certain disk models of X-ray
binaries, 
in particular when 
the disk is assumed optically thick and its 
stresses proportional to the radiation pressure \citep{LE74,SS76}. 
But there has always been a
question mark regarding the applicability of viscous instability to
real accretion disks because of the last assumption. It is by no
means assured that
 the turbulent stresses in
radiatively dominated accretion flows behave in the way required \citep[but see][]{Hirose09}.

Inspired by the \emph{Voyager} images, 
early theories of planetary ring structure appealed to the
viscous instability \citep{Ward81,Lukkari81,LB81},
but again doubts were raised about
the properties of the viscous stress, and interest dwindled. Kinetic theory indicates that
dilute rings possess a stress
that decreases with surface density \citep{GoldTrem78,SS85}
 but also that 
dense rings emphatically do not \citep{ArakiTrem86}.
If viscous instability is to occur in dense rings, it must attack only
the smallest particles selectively \citep{SaloSchmidt10}.

\subsection{Thermal instability and `phase' changes}

Gaseous accretion disks may fall into one of many thermal/ionisation
equilibria for a given set of parameters. Not all of these are
thermally stable and so it is possible that a disk cycles
between different states over time, generating
quasi-periodic variability in accretion 
luminosity and non-thermal emission. 

\begin{figure}
\figurebox{18pc}{}{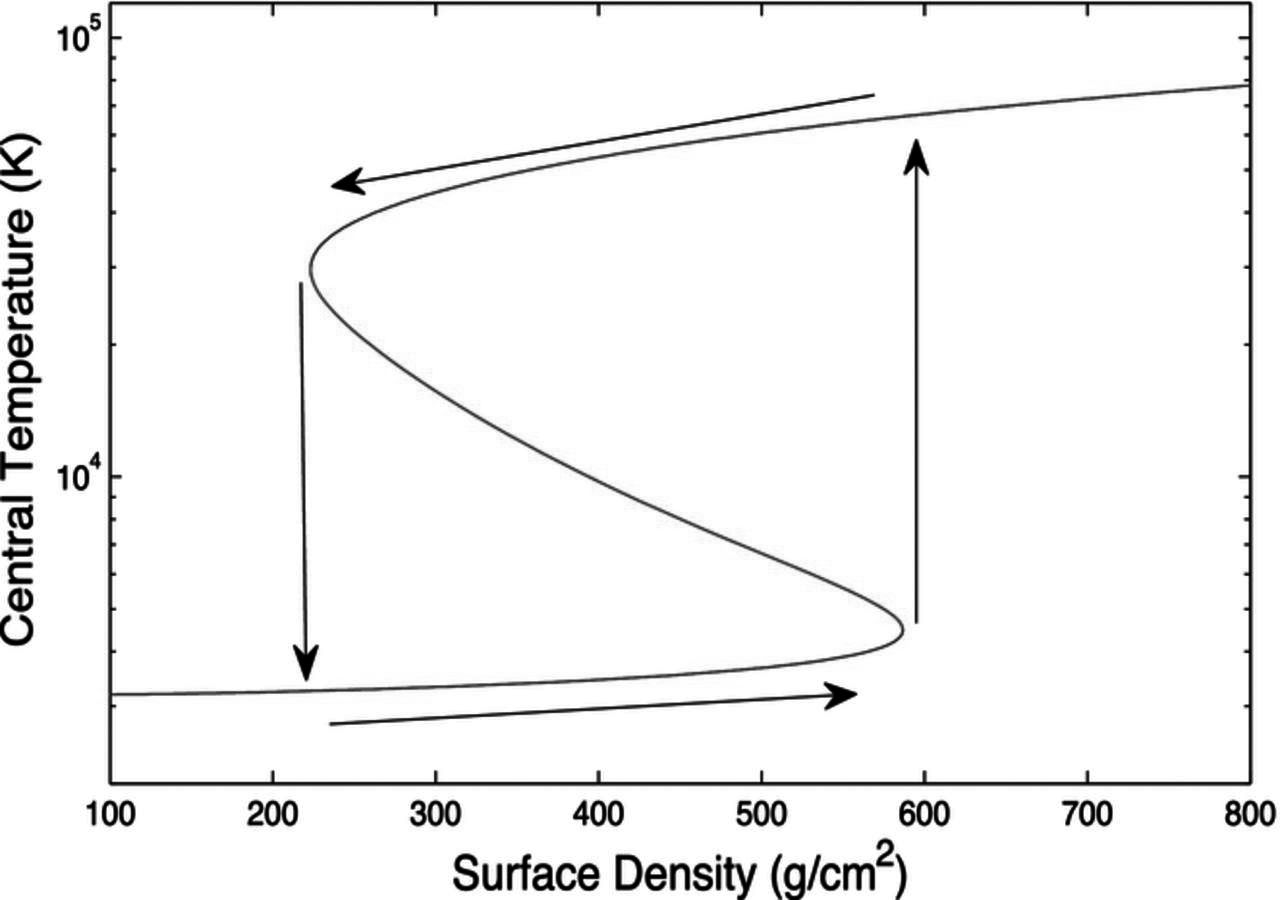}
\caption{A characteristic dwarf nova S-curve, describing
  thermal equilibria in the phase space of surface density and
  temperature. The arrows indicate a limit cycle that the disk may follow.
Credit: \citet{LatPap12}. \label{fig:Scurve}}
\end{figure}

The classic and best understood
examples are dwarf novae which straddle the temperature threshold for
hydrogen ionisation (about 5000 K). Because the opacity
increases dramatically in the partially ionised phase, the gas's cooling rate
is a complicated function of the temperature and permits the
disk to
support three possible thermal equilibria for a given surface density
\citep{MMH81,Faulk83}.
In the phase space of surface density and temperature, disk equilibria
sketch out a characteristic `S-curve', an example of which we plot in
Figure \ref{fig:Scurve}. The disk may then enter a limit
cycle, oscillating between the hot, well-ionised, and luminous high
state (an outburst) and the cool, poorly ionised, and dim low state. The transition
between the two states takes place via thermal fronts that rapidly
sweep through the disk. The story is complicated by a raft of ancillary physics, but
in general the contact with observations is relatively good. Similar
physics is shared with low-mass X-ray binaries, but these systems
are not so well understood and the model enjoys less success \citep{Lasota01}.

Disks dominated by radiation pressure 
can support a variety of
interesting 
equilibria partly because of the relative inefficiency of cooling in very hot
plasma. In addition to the standard thin disk solution of \citet{SS73}, 
there exist `slim' and `thick' disks in which
radiative cooling is supplanted by radial advection of energy by the
gas \citep{Abramo88}. The most extreme case is the advection-dominated
accretion flow (ADAF) where little of the dissipated energy
is radiated locally and the rotation profile deviates significantly
from Keplerian \citep{NY94}. If the turbulent
stresses are assumed proportional to the total pressure, these solutions are
thermally unstable because the heating depends on temperature to a
much greater power than cooling \citep{SS76,Piran78}. 
Even though recent MRI simulations indicate that thermal
instability can arise in such disks \citep{JS13}, 
X-ray observations of strongly accreting black hole systems
fail to exhibit unstable or cyclic dynamics on the expected timescales
\citep{GD04,DGK07}. 
Only the exceptionally luminous black hole binary, GRS 1915+105, shows
anything like an expected limit cycle driven by thermal instability
 \citep{Done04}.
In contrast there are abundant observations of other kinds 
of variability, especially in the
spectroscopy, and a panoply of well-defined
 states exist in the phase space of intensity,
hardness, and rms fluctuation \citep{RemillardMcClintock2006,Belloni10}. 
To date there is no convincing explanation for the cycling
between these phases.

Finally, protoplanetary disks are thought to jump aperiodically 
between high and low accreting states on a time scale of 100+
years. This outbursting
behaviour is exemplified by the archetype FU Orionis, though other classes
may exist, notably the shorter timescale EXors \citep{Audard2014}.
The current theoretical paradigm posits that the low state
corresponds to a disk whose central region (the `dead zone') is cold, 
laminar and inefficiently accreting and that the
high state corresponds to one in which this region is engulfed in MRI
turbulence. The build-up of mass at the outer edge of the dead zone and
a consequent
gravitational instability are the trigger for an outburst 
\citep{Gammie96,Armitage01}.

Saturn's rings also exhibit multiple phases.
In the inner B-ring there exist adjacent flat and undulatory states,
while the middle and outer B-ring breaks up into disjunct bands of
intermediate and high optical depth; 
both phenomena occur on scales of order 100 km 
\citep{Colwelletal09}. In
accretion disks the different phases are distributed over
time, but in planetary rings, with their achingly slow timescales,
adjoining states are distributed spatially. For example, during a dwarf nova outburst
the high state overwhelms the disk in $\sim 1$ day; in
contrast, the undulatory state in the inner
B-ring may take over the entire ring in $10^{10}$ days \citep{Latter12}.

The origin of the observed states in planetary rings is not well understood. It is
perhaps unlikely that they issue from a bistability in the thermal
equilibrium itself (as in dwarf novae).
For a constant coefficient of restitution $\epsilon$, there
is only ever one equilibrium for a given set of 
parameters \citep{ArakiTrem86}, while experimentally derived laws describing
$\epsilon$'s dependence on impact speed yield the same result \citep{WisTrem88,Salo91,LO08}. 
Note, however, that
 if $\epsilon$ is a non-monotonic function
of impact speed then bistability could be possible. 
Such a coefficient of restitution may correspond to collision models
that incorporate particle sticking at low impact speeds \citep{AlbersSpahn06}.

The undulatory and flat states in the inner B-ring could be
competing outcomes of the ballistic transport instability 
\citep{Durisen95}, whose nonlinear dynamics supports bistability 
\citep{Latter14a,Latter14b}. The 100 km structures deeper in the B-ring are more
mysterious. \citet{Tremaine03} proposed that these correspond to shearing and
shear-free regions, the latter held together by strong inter-particle
adhesion. While it may be possible to hold together 100 km shear-free
bands 
if the disk were restricted to the orbital plane, in three-dimensions
the rings' tensile strength will be too weak because the rings can vertically `relax'.

\section{SATELLITE--DISK INTERACTIONS}
\label{SDinteractions}
\subsection{Introduction}

Astrophysical bodies that are surrounded by continuous disks often
possess discrete satellites as well.  Examples include the moons of giant
planets, which then interact with their planetary rings, protoplanets interacting
with a circumstellar disk, and binary stars and black holes
interacting with accretion disks.  Various geometrical configurations
are possible, as a satellite can orbit fully interior or exterior to
the disk, be confined within an annular gap, or be fully embedded in
the disk.

The gravitational interaction between an orbital companion and a disk
is a problem of general interest in astrophysics.  By generating waves
and other disturbances in the disk, the satellite both undergoes orbital
evolution and influences disk properties.
In addition, observations of the structures induced in the disk can be used to
constrain physical properties of both the disk and the perturber.

One important application, discussed in greater detail below, is to
planets formed in a gaseous disk around a young star.  The
planet--disk interaction causes the planet to migrate radially through
the disk, a process that needs to be understood and quantified in
order to interpret the observed properties and architectures of
exoplanetary systems (as well as the Solar System).  Ever since the
first planets were discovered on orbits very close to solar-type
stars, it was suggested that planetary migration brought them to their
current locations \citep{MayorQueloz1995}.  As more and more hot
Jupiters have been found, it is generally accepted that these
planets formed at locations beyond~1~AU~and have since migrated
inwards.

Orbital migration can also be important in AGN.  When two galaxies
merge, the central black hole of the smaller galaxy interacts with the
accretion disk surrounding the larger black hole, in a way analogous
to a planet interacting with a circumstellar disk.  Inward orbital
migration leads eventually to a compact binary black hole that merges
as a result of gravitational radiation, as spectacularly confirmed by
recent observations \citep{2016PhRvL.116f1102A}.

Finally, many of the observed structures in astrophysical disks and planetary
rings can be attributed to satellite--disk interactions.  Examples
include the Cassini division, the Encke and Keeler gaps, the spiral
waves in Saturn's rings, as well as the
 spiral arms in interacting galaxies and the
tidal truncation of accretion disks in binary stars.  Planet--disk
interaction can also create annular gaps and spiral structures in PP
disks, some examples of which may have been recently observed (see
section~\ref{s:PPdisks}).

\subsection{Wave launching, coorbital torques and type-I migration}

The simplest situation involves a satellite on a circular orbit that
is coplanar with the disk.  The satellite exerts a periodic
gravitational force on the disk and excites its epicyclic motion.  The
forced motion is resonant at a series of radii, located both interior
and exterior to the satellite's orbit, and the disk responds by
launching a spiral density wave at each of these resonances.  To the
extent that the epicyclic frequency corresponds to the orbital
frequency, these Lindblad resonances can be identified with
mean-motion resonances involving commensurabilities of the form
$m:m\pm1$ between the orbital frequencies of the satellite and the
disk; in fact they are slightly shifted radially because of small
departures from Keplerian motion due to pressure gradients, disk self-gravity, oblateness
of the central body, etc.

The act of launching a wave involves a transfer of energy and angular
momentum from the satellite's orbit.  As the waves propagate radially
away from the Lindblad resonances, their radial wavelength decreases
and they are damped by viscosity or other dissipative processes.
The damping may be enhanced if the waves attain nonlinear amplitudes.
The angular-momentum flux carried by the wave is transferred to the
disk as the wave is damped and thus a resonant torque is exerted between
the satellite and the disk.

Numerous examples of these density waves in Saturn's A-ring were
observed by the \emph{Voyager} and \emph{Cassini} spacecraft, and have
been identified with specific Lindblad resonances with various moons
that orbit outside the A-ring.  A different type of density wave is
seen near the edges of the Encke and Keeler gaps in the outer A-ring.
These are excited by the satellites that orbit within these gaps and
are therefore closer to the ring material than the larger external
moons.  In these cases the wake cannot be identified with a single
Lindblad resonance, although it can be thought of as a superposition
of waves generated at many such resonances of high order.

\begin{figure}
\figurebox{19pc}{}{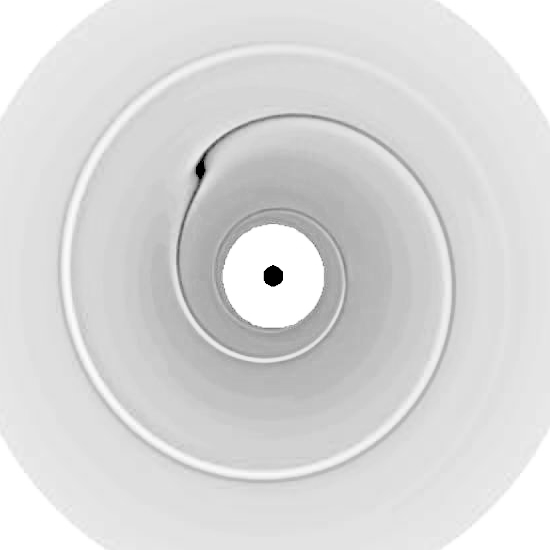}
\caption{Hydrodynamic simulation of a planet of 10 Earth masses
  undergoing type-I migration in a protoplanetary disk. The surface
  density is plotted in grey-scale.}  \label{fig:typeI}
\end{figure}

In the case of planets interacting with PP disks, related phenomena
have manifested chiefly in theoretical work \citep[although there is
now some observational evidence of spiral waves in PP disks, which can
be explained best by embedded
planets;][]{2015ApJ...809L...5D,2016ApJ...816L..12D}, and 
in recent years
hydrodynamic simulations have led the way in determining the nonlinear
dynamics of
planet--disk interactions. Embedded satellites that are not massive
enough to open a gap in the disk's density profile undergo what is
called \emph{type-I migration}
\citep{GoldreichTremaine1978b,GT80,1997Icar..126..261W}.  The
density waves launched at different Lindblad resonances constructively
interfere to produce a coherent one-armed spiral wake
\citep{2002MNRAS.330..950O}, a narrow overdense region.  The
wake is not symmetric about the satellite's location, because of the
circular geometry and possible radial gradients in the properties of
the unperturbed disk.  The satellite therefore experiences a net
gravitational torque, which under most circumstances is negative,
leading to inward migration of the satellite.
 A snapshot of the
surface density in a hydrodynamic disk simulation with an embedded
planet of 10 Earth masses is shown in Figure~\ref{fig:typeI}


This description is incomplete because a different type of interaction
happens closer to the satellite's orbit, where disk material
approximately corotates with the satellite.  In this region the
relative motion of the disk and satellite is too slow for density
waves to be excited, but the satellite can instead generate
non-wavelike disturbances in the potential vorticity and entropy of
the disk, each of which involves an asymmetric rearrangement of the
surface density and therefore a torque.  In distinction to the Lindblad
torques associated with the launching of density waves, the torques
arising from this region are known as coorbital, corotation or
horseshoe torques.  
The last name comes from the property that, in
the frame rotating with the satellite's orbit, disk material in the
coorbital region has streamlines that librate rather than circulate,
and involve horseshoe-shaped turns near the satellite's longitude.


The entropy-related corotation torque can lead to stalling
of planets exceeding roughly 3-5 earth masses
at specific disk locations that vary with time \citep{B14}. Ultimately,
however, these torques are
sustained by dissipative effects, such as viscosity, thermal
diffusivity, and radiative cooling which are all uncertain 
\citep{2008A&A...485..877P,2009MNRAS.394.2283P}.
In addition, the physics of the coorbital region is strongly
nonlinear, and thus difficult to describe accurately.
The net (Lindblad
plus coorbital) torque is subject to similar uncertainties.

Nevertheless, there is general agreement that planets of Earth mass
typically migrate towards the star on a timescale of a few hundred
thousand years.  This poses a problem for planet formation because the
viscous timescale, i.e.\ the lifetime of the accretion disk, is
thought to be significantly longer than that. The inconvenient
conclusion is that every Earth-mass
planet in a protoplanetary disk should have migrated into the star.
Various ways to prevent this from happening have been proposed.  These
ideas include the stochastic torques arising from turbulence in the
disk (see later), positive torques associated with asymmetric heating
in the planet's neighbourhood \citep{BL15}, and
so-called \emph{planet traps} due to dead zones, snow
lines or other features in the disk \citep{Masset06}. Similar physics may also control the
migration of propellers in Saturn's rings \citep{Tisc13}.
We still, however, lack a general theory
capable of predicting the speed and direction of type-I migration,
which is vitally needed to link together the early and late stages of
planet formation.

\subsection{Gap opening and type-II migration}

In the type-I regime, while the fractional surface density
perturbation in the wake can become of order unity at some distances
from the satellite (involving a shock, in the case of a gas disk), the
azimuthally averaged surface density of the disk is not significantly
perturbed.  More massive satellites, however, are able to deplete the
disk locally by opening an annular gap.  This happens because the
Lindblad torques exerted by the satellite on the interior and exterior
parts of the disk are negative and positive, respectively, causing
both to recede from the satellite's orbit.

The scaling laws for gap opening have been formulated by
\citet{LinPap86,LinPap93}. Note that it is difficult to quantify the
process exactly because gaps can be of differing degrees of
cleanliness. In order to open a gap, the mass ratio $M_\mathrm{s}/M$
of the satellite and the central object should be sufficiently
large. A first criterion is that $M_\mathrm{s}/M\gtrsim(H/r)^3$, or,
equivalently, that the Hill radius $R_\mathrm{H}\gtrsim H$. This is
the condition for the disturbance generated in the disk to be
nonlinear close to the satellite, which allows it to be dissipated
locally. A second criterion, which is usually more stringent, is that
$M_\mathrm{s}/M\gtrsim(81\pi/8)(\nu/r^2\Omega)$. This is the condition
that the tidal torque exceed the viscous torque in the undisturbed
disk (if indeed it can be adequately described by a simple kinematic
viscosity).  
When applied to a protoplanetary disk, the latter criterion implies
that a planet of about Saturn's mass or greater is able to open a gap.
When applied to the outer part of Saturn's A-ring, the criteria imply that a
moon of about Daphnis's mass or greater is able to open a gap.

\begin{figure}
\figurebox{19pc}{}{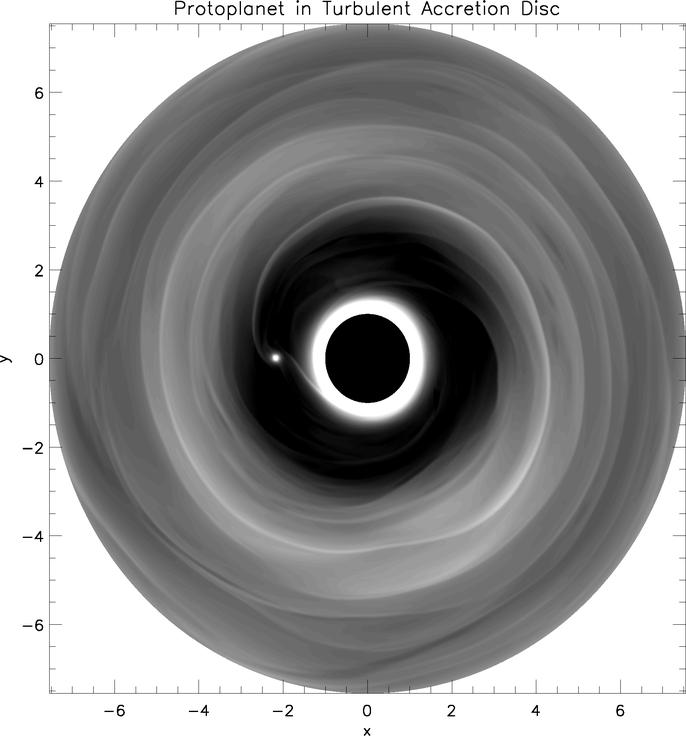}
\caption{Hydrodynamic simulation of a planet of 5 Jupiter masses
  opening a gap and undergoing type-II migration in a turbulent
  protoplanetary disk. Again the surface density is plotted.
  Credit: Nelson \& Papaloizou (2003).}  \label{fig:typeII}
\end{figure}

The characteristics of satellite migration change significantly when a
gap opens --- a situation referred to as \emph{type-II migration}
(Figure~\ref{fig:typeII}).  The gap effectively divides the disk into
interior and exterior disks, making it difficult (or impossible) for
material to pass through.  The satellite is effectively locked to the
viscous evolution of the disk, and may even retard that evolution if
it is more massive than the interior disk.  The timescales of type-II
migration are therefore in general longer than those of type-I
migration.  HL~Tau (see Figure~\ref{fig:hltau}) might be a system
where planets were able to open gaps, although confirmation of this
interpretation is still pending.

\subsection{Stochastic migration}

Classical theories of type-I and type-II migration assume a smooth,
laminar disk as a background state.  Since disks are typically active
with a number of instabilities, this may be a poor approximation.
Instabilities can lead to turbulence and cause density perturbations
on many scales.  This can, especially for small embedded satellites,
lead to migration that resembles a random walk in the orbital
parameters \citep{NelsonPapaloizou04,ReinPapaloizou2009}.  This
\emph{stochastic migration} is relevant for low-mass planets in PP
disks in the presence of the MRI or GI, and for moonlets in planetary
rings (discussed in detail in the Spahn chapter).
If the moonlets are small enough, their migration will be dominated by
the stochastic component \citep{ReinPapaloizou2010}, rather than the
classical type-I torque.  The main contributions to the stochastic
migration of moonlets in rings are collisions with ring particles
(which have no analogue in PP disks) and interactions with the
(temporary) clumps and overdensities created
predominantly by self-gravity wakes. 

\subsection{Tidal truncation and disk edges}

Satellites that are significantly more massive than those that are
able to open a gap can truncate a disk at great distance.  In Saturn's
rings the outer edge of the B-ring is associated with the 2:1
Lindblad resonance with Mimas, while the outer edge of the A-ring
coincides with the 7:6 resonance with the coorbital moons Janus and
Epimetheus.  Related phenomena occur in accretion disks in binary
stars; for typical mass ratios, however, the 2:1 resonance is so
strong that the disk is truncated well inside it \citep{PapPri77,Pac77}.
On the other hand, the $\epsilon$-ring of Uranus appears to be
`shepherded', i.e.\ radially confined, by the
satellites Cordelia and Ophelia on either side, via discrete
Lindblad resonances \citep{GolPor87}.

In order to explain the extreme sharpness of the edges of planetary
rings, in particular those mentioned above, it has been found
necessary to appeal to a modification of the viscous torque that
occurs when the ring carries a nonlinear density wave
\citep[][Chapters by Longaretti and Nicholson]{BGT82,BGT89}.  
The presence of the wave alters the velocity
gradient associated with circular orbits in such a way that the
angular-momentum flux is reduced to zero when the wave has a critical
amplitude, and is reversed for waves of higher amplitude.  Without
this effect, the edges of planetary rings would be smoothed out by
viscosity and it would be difficult or impossible to account for the
shepherding process.  It is not known whether the modification or
reversal of the angular-momentum flux plays a role in gaseous disks.

\subsection{Eccentricity and inclination}

If the satellite's orbit is eccentric or inclined with respect to the
disk, then the gravitational force on the disk contains additional
components that launch density or bending waves at Lindblad or
vertical resonances.  The eccentricity or inclination of the satellite
evolves as a result of these interactions.  For satellites embedded in
the disk, a small ($\lesssim H/r$) eccentricity $e$ or inclination $i$
is
damped on a timescale that is shorter than that for type-I migration.
This interaction is dominated by disk material close to the planet.
For larger $e$ or $i$ the damping is less efficient and the direction
and rate of migration may also be affected.  When the planet's speed
relative to the local gas exceeds the sound speed or velocity
dispersion (as happens for example when the inclination of the
satellite brings it above the disk, i.e.\ $i\gtrsim H/r$) the
interaction is dominated by gas drag similar to that of a star passing
through the interstellar medium \citep{Rein2012}.

More massive satellites that are well separated from the disk material
interact through more distant orbital resonances and the net effect
can, under some circumstances, be a growth in $e$ and/or $i$
\citep{GT81,BGT83,BGT84}.  In fact, the disk can also become elliptical
and/or warped through these interactions, and the relative magnitude
of the $e$ or $i$ of the disk and the satellite depends on their
coupled dynamics \citep{LubOgi01,TeyOgi16}.

A good example of satellite--disk interaction producing a growth of
eccentricity occurs in SU Ursae Majoris binary stars, where the accretion disk
around a white dwarf is understood to become elliptical as a result of
an interaction with the companion star at the 3:1~orbital resonance,
an example of an eccentric Lindblad resonance \citep{Lubow91}. This is
the `superhump' phenomenon, so called because of the associated
modulation of the light curve during a superoutburst.

A similar process may be responsible for maintaining the eccentricity
of Uranus's $\epsilon$-ring; the 47:49 eccentric Lindblad resonance
with its inner `shepherd' satellite, Cordelia, lies within the ring
\citep{GolPor87}.  However, it should be noted that other
narrow rings around Saturn and Uranus are also found to be eccentric
even though they have no observed shepherds (see Nicholson chapter).

\section{CONCLUSION}

In this chapter we make explicit connections
between the study of planetary rings and of other astrophysical
disks, putting an emphasis on dynamics.  
Disk systems exhibit an enormous physical
and dynamical diversity, but one anchored upon the fundamental balance between
radial gravity and the centrifugal force.  A relic of formation,
 their orbital angular momentum is inherited from the collapse of a
cloud, the disruption of body by a massive companion, or the collision
of two bodies around a more massive object. This `excess' angular momentum
thwarts the simple accretion of disk material upon the central body,
irrespective of the different formation routes, and leads to planetary
rings and astrophysical disks.

Through the diversity of composition and physics
one can discern recurrent themes. To begin, most gaseous disks support
hydrodynamical (or magnetohydrodynamical) activity that permits
disk material to slowly shed or redistribute its angular momentum
 and thus ultimately accrete. By liberating
orbital energy this activity also causes the disk to radiate
 --- the key to understanding certain high energy sources.
The outward transport of angular momentum (whether mediated
by turbulent motions in gas or collisions between particles) 
controls the evolution and lifetime of the disk
or ring. Though the `viscous' lifespans of
planetary rings and gaseous disks differ by orders of magnitude, 
accretion represents one of their key
connections.

Collisional dynamics presents a
link between rings and other particulate disks, such
as debris disks and the belts of dust and planetesimals orbiting
young stars. 
In the ring context, collisions not
only help transport angular momentum but also control the
composition, structure, and size evolution of the constituent
particles themselves. The same processes govern PP and
debris disks,
leading to the formation of planetesimals and
planets in the former, and dust in the latter.

Both rings and disks support the passage of waves and the growth of
instabilities, which contribute to the activity required to sustain
accretion and angular-momentum transport.  Regarding self-excited
instabilities, only in the case of gravitational instability
 is (nearly) the same process
reliably occurring in both rings and gaseous disks, although viscous
overstability may be a second example.  
Satellite--disk interactions,
on the other hand, provide the richest set of dynamics shared by the two classes
of systems.  Spiral density waves, gap formation, and satellite
migration, all now directly observed in detail around Saturn,
have important analogues around young 
stars that are beginning to yield observational manifestations (with
instruments such as ALMA).
In that respect, observations of satellite--disk interactions 
in planetary rings are ahead of those in PP disks by a few decades.
But just as the Voyager data proved so exciting and fruitful during the
1980s, so should ALMA observations in the following decade, as its
full capabilities
are brought to bear on the problem of exoplanets and their host disks.     

Given these overlaps, it is no surprise the fields of disks and of rings
have connected in mutually beneficial ways. 
The clearest instance
is in the study
of satellite--disk interactions, where work on binary
stars and Saturn's rings converged fruitfully in the theoretical
understanding of 
planet--disk coupling in the 1980s. Another example is the research in
disk instabilities, first introduced in gaseous disks in the 1970s
but matured in the planetary ring context over the next few decades. 
 Despite this
 historically close connection, there yet remain a number of
correspondences that have yet to be fully
capitalised upon and which form the basis for several
appealing research directions. 

The size-distribution dynamics of dense rings is
an underdeveloped area compared with that for debris disks and planet formation. 
Though arguably a more difficult problem, 
the techniques and tools of the latter could be profitably adapted
to help explain the
distributions in Saturn's rings, which are
observationally well constrained in comparison to those more distant
particulate systems.
Note that the
 statistical approaches used in debris disks may be applied without too much complication
 to the F-ring, as it is probably the closest ring analogue.
On the other hand, this is also an opportunity to determine
  how well these techniques and tools perform on an object so much better
  constrained than a debris disk.
 An especially compelling
sub-question is
the provenance of the propellers in the A-ring. Is it possible that these large objects
are related to the `lucky' 
planetesimals that grow to large sizes in PP disks, their
smaller brethren languishing in the cm to m size classes? 

Another area of fruitful overlap is in the detailed microphysics of 
collisions. Only recently have studies of dense planetary rings
 moved away
from the bouncing hard-sphere model of ring particles. 
In contrast, the numerical treatment of collisions in
planet formation is much better developed, allowing for the full gamut
of physical processes (compaction, mass transfer, fragmentation,
reaccretion, etc.). 
Is it possible to establish clear barriers to growth in dense rings, as in PP
disks? Can we construct a typology
of collisional outcomes in dense rings as clearly as in 
planetesimal belts? 

Gravitational
instability (and gravitoturbulence) is shared by planetary rings, PP
disks, and galactic disks. However, the details of its onset and saturation are still
unclear in each context.
A more unified approach would
nail down its manifestation in all three classes, and
indeed uncover more profound connections with other subcritical
transitions to turbulence in shearing and rotating systems.
Another area worth exploring is the production of tidal streams 
around white dwarfs, which clearly shares the same physics of certain ring formation
scenarios, especially of narrow rings.
Finally, the field of satellite--disk interactions, though mature, could
but undoubtedly support further connections between rings and disks. 
As observations of PP disks become more and
more detailed due to ALMA, the intricate and
varied morphologies supported by Saturn's rings will provide valuable
analogues with which to understand them.

\section*{Acknowledgments}

The authors thank the anonymous reviewer and the editors Matthew
Tiscareno and Carl Murray for a set of helpful
comments. They are particularly indebted to the generosity of
colleagues and friends 
who read through earlier stages of the manuscript, in
particular Julia Forman, Cleo Loi, Pierre-Yves Longaretti, John
Papaloizou, and
Mark Wyatt. The chapter was much improved by their insightful and
helpful remarks. We also thank Tobias Heinemann and Antoine
Riols-Fonclare for providing figures.

\bibliography{cplslargesample}\label{refs}
\bibliographystyle{cambridgeauthordate}

 \printindex

\end{document}